\documentclass[12pt,fleqn]{article}
\usepackage{espcrc1}
\readRCS
$Id: espcrc2.tex,v 1.2 2004/02/24 11:22:11 spepping Exp $
\usepackage{epsfig}
\usepackage{amsmath}
\usepackage{amssymb}
\usepackage{euscript}
\usepackage{color}
\title{{\bf Luminosity Measurement Method for the LHC: \\
Event Selection and Absolute Luminosity Determination}\thanks{This work was 
supported in part by the programme of co-operation between the IN2P3 and 
Polish Laboratories No. 05-117, Polonium Programme No. 17783NY and 
Polish Grant No. 665/N-CERN-ATLAS/2010/0.
}}
\author{ M.~W. Krasny\address[LPNHE]{ LPNHE, Pierre and Marie Curie 
                         University, CNRS-IN2P3,  Tour 33, RdC, \\
                         4, pl. Jussieu, 75005 Paris, France.}, %
J. Chwastowski\address[PK]{ Institute of Teleinformatics, 
Faculty of Physics, Mathematics and Computer Science,
        Cracow University of Technology,
        ul. Warszawska 24, 31-115 Krak\'ow, Poland.}
\address[IFJ]{ Institute of Nuclear Physics PAN,
ul. Radzikowskiego 152,
31-342 Krak\'ow, Poland.}, %
A. Cyz\addressmark[IFJ] %
and
K. S{\l}owikowski\addressmark[IFJ]}

\begin{document}

\maketitle
\begin{abstract}
\vspace{1cm}
Absolute normalisation of the LHC measurements with $\cal{O}$(1\%) precision
and their relative normalisation, for the data collected at variable 
centre-of-mass energies, or for variable beam particle species, with  
$\cal{O}$(0.1\%) precision is crucial for the LHC experimental programme but
presently beyond the reach for the general purpose LHC detectors. This paper 
is the third in the series of papers presenting the measurement method  capable 
to achieve such a goal. 
\vspace{1cm}
\end{abstract}

\section{Introduction}
\label{sec:Introduction}

The requirements for the luminosity measurement precision at the LHC are often 
misunderstood. According to the present paradigm \cite{Mangano}, 2\% precision 
is the benchmark and the ultimate target for the LHC experiments. Such a target 
may soon be reached by the  method based on van der Meer scans for which  
$\delta L /L = \pm 3.7~\%$  has already been achieved \cite{vdM}.  

The main argument underlying such a paradigm reflects the present precision of 
theoretical calculations of the cross sections for hard parton processes. These 
calculations are based on the perturbative, leading-twist QCD framework which
allows only to approach the experimental precision target. Thus, as long as a 
significant reduction of uncertainties of both the matrix elements and the 
parton distribution functions (PDFs) is beyond the reach of the available 
calculation methods and DIS experiments, the impact of further reduction of the 
experimental luminosity error would hardly improve the present understanding of 
parton collisions.

In our view, there are at least four reasons to push the luminosity measurement 
precision frontier as much as possible. 
\begin{enumerate}

\item Precision observables, based on the data collected at variable 
c.m.-energies  have been  proposed for the precision measurement programme of 
the Standard Model parameters at the LHC \cite{krasnySMparameters}, and for the 
experimental discrimination of the Higgs production processes against the SM 
ones \cite{krasnyHiggs}. These observables are specially designed to drastically 
reduce their sensitivities to the theoretical and modeling uncertainties 
limiting the overall measurement precision of the canonical ones. Their 
measurement precision is no longer limited by the theoretical and modeling 
uncertainties but by the precision of the relative luminosity measurements at 
each of the collider energy settings.

\item The general purpose LHC experiments, ATLAS and CMS, and the specialised 
experiment, LHCb, measure the same parton cross sections in different 
regions of the phase-space. Relative  normalisation of their results  
determine the ultimate LHC precision for those of  the measurements which are  
based on a coherent analysis of their data. The best example here is the 
measurement of $\sin^2(\theta _W)$ associated with concurrent unfolding of the 
valence and sea quarks PDFs.   

\item The LHC collides proton beams while the Tevatron collides proton and 
anti-proton beams. The $W$ and $Z$ boson differential cross sections measured at 
these two colliders constrain the flavour dependent PDFs within the precision of 
the relative LHC and Tevatron luminosity measurements because the theoretical 
uncertainties on the hard partonic  cross sections largely cancel in the cross
section ratios.

\item A fraction of the LHC running time is  devoted to collisions of ions. The 
precision of the relative normalisation of the $pp$ interaction observables with 
respect to the corresponding ion-ion ones is  crucial for a broad spectrum of 
measurements ranging from the classical measurements of the shadowing effects to 
more sophisticated ones including the experimental studies of the propagation of 
W and Z bosons in the vacuum and in the hadronic matter \cite{krasny_jadach}. 

\end{enumerate}  
\noindent
For all the above and many other reasons any future improvement of the 
luminosity measurement precision will be directly reflected in the increased 
precision of the interpretation of the LHC measurements in terms of the Standard 
Model (SM) and Beyond the Standard Model (BSM) processes, in particular if 
specially designed precision observables are  used.  

The method of the luminosity measurement proposed in this series of papers 
attempts to push the precision frontier as much as  possible at hadron 
colliders. Its ultimate goal is to reach the precision of $\cal{O}$(1\%) for the 
absolute normalisation of the LHC measurements and of $\cal{O}$(0.1\%) 
for the relative normalisation of the data recorded at variable centre-of-mass 
energies or different beam particle species (protons, ions). It is based on the 
detection of those of the electromagnetic collisions of the beam particles in 
which they can be treated as point-like leptons. For these collisions the 
corresponding cross sections can be calculated with a precision approaching 
the one achieved at lepton colliders. In this series of papers such collisions  
are tagged by the associated production of the unlike-charge electron-positron pairs. 

In the initial paper \cite{first} of this series we have selected the  
phase-space region where the lepton pair production cross section is 
theoretically controlled with precision better than 1\%, is large enough to 
reach a comparable statistical accuracy of the absolute luminosity measurement 
on the day-by-day basis and, last but not the least, its measurement 
is independent of the beam emittance and of  the Interaction Point (IP) optics. 
Collisions of the beam particles producing lepton pairs 
in this phase-space region cannot be triggered and selected by the present LHC 
detectors. A new detector must be incorporated within one of the existing 
general purpose LHC detectors to achieve this goal.

 In our second paper \cite{second} the performance requirements for a dedicated 
 luminosity detector were discussed. In the present paper we discuss the luminosity 
 measurement procedure based on the concrete luminosity detector model and on 
 the present  host detector performance parameters. We evaluate quantitatively the systematic 
precision of the proposed luminosity measurement method. 

This paper is organised as follows. In Section 2 the model of the luminosity 
detector is introduced. Section 3 specifies the performance requirements for the 
host detector. The on-line selection procedures for the luminosity events 
are proposed  in Section 4 and optimised in Section 5. Section 6 presents the 
rates of the signal and background events at the consecutive stages of the event 
selection process. The methods of monitoring of the instantaneous, relative 
luminosity are described in Section 7. Section 8 introduces the luminosity 
measurement methods for the low, medium and high luminosity running periods of 
the LHC machine. Merits of the dedicated runs are discussed in section 9. 
Section 10 introduces a novel and simple method of calculating the absolute cross sections 
for any sample of the user-selected events. Finally, Section 11 is devoted to 
the evaluation of the systematic uncertainties of the proposed luminosity 
measurement method.

\section{Luminosity Detector Model}

In the studies presented below the model of the luminosity detector is specified 
in terms of a minimal set of its required output signals. Its concrete hardware 
design is open and depends solely on the specific constraints imposed by the 
host detector construction and by the wishes of the host detector collaboration. 
The luminosity detector can be realised using one of the available particle 
tracking technologies and does not require dedicated R\&D studies. The D0 fibre 
tracker \cite{D0tracker} can serve as an example in most of its functional 
aspects with a notable exception of the radiation hardness issue which would 
need to be addressed anew in the LHC context.    

The geometry of the proposed luminosity detector model was presented in 
\cite{second}. The 
detector fiducial volume consists of two identical cylinders placed 
symmetrically with respect to the beam collision  point. Each cylinder is 
concentric with the proton-proton collision axis defined in the following as the 
$z$-axis of the coordinate system. The cylinders have the following dimensions: 
the inner radius of 48~cm, the outer one of 106~cm and the length of 54.3~cm. 
They occupy the regions between $z_f^{right} = 284.9$~cm and  $z_r ^{right}= 
339.2$~cm and $z_f ^{left}= -284.9$~cm and  $z_r ^{left}= -339.2$~cm. Each 
cylinder contains three layers (the $z_1, z_2, z_3$ planes) providing  the 
measurements of the charged particles hits. These planes are positioned at the 
distances of $z_1 = 285.8$, $z_2 = 312.05$ and $z_3 = 338.3$~cm from the 
interaction point. Each of the luminosity detector $z$-planes is segmented into 
3142 $\phi$-sectors providing the azimuthal position of the hits\footnote{Adding 
the pseudorapidity segmented $z$-planes (or equivalently $\phi$-tilted planes) 
would certainly be useful, in particular for the periods of the highest LHC 
luminosities. Since, the main purpose of this paper is a proof of principle, 
these aspect will not be discussed further here. The presented luminosity 
detector model can be thus considered as the model of the base-line detector 
satisfying a minimal set of requirements.}.
Such a detector could occupy the space foreseen for the TRT C-wheels - a 
sub-component of the ATLAS detector which has not been built. 

The basic role of the luminosity detector is to provide, for each charged 
particle $j$, produced in $pp$ collisions, the ``track segment", specified in 
terms of the three azimuthal hit positions, $i_{j,k}$, where $k \in \{1, 2, 3\}$ 
is the plane number and $j$ denotes the particle number. It is assumed that each  
reconstructed track segment has a ``time-stamp". The time-stamp assigns the 
track segment to the ${\cal O}(1)$~ns wide time slot synchronised with the 
machine clock\footnote{Note that this requirement may lead to adding, if 
necessary for a chosen technology, an extra, coarse $\phi$-granularity 
timing plane in front of the three highly segmented hit planes.}. 
Further timing requirements, discussed in detail in \cite{second}, such as  
adding a precise, sub-nanosecond relative timing for each of the $\phi$-hits, 
for a better level 1 (LVL1) trigger resolution of the $z$ position of the origin 
of the track segments, are not required in the base-line detector model. 

The time stamped track segments and their associated hits are the input data for 
the algorithmic procedures allowing to select, within the 2.5$\mu s$ LVL1 
trigger latency of the host detector, the ``luminosity measurement 
bunch crossings" and ``monitoring bunch crossings". The subsequent event 
selection algorithms are based solely on the host detector signals. The ``Region 
of Interest" (ROI) \cite{ATLAS} mechanism is used to correlate, within the level 
2 (LVL2) trigger latency, the luminosity detector data with the relevant data 
coming from the host detector.

\section{Host Detector Model}

The ATLAS detector \cite{ATLAS} was chosen in our studies as the host detector. 
The following data coming from this detector are used in the LVL2 and Event 
Filter (EF) trigger algorithms:  

\begin{itemize} 
\item the parameters of the tracks selected using the ROI mechanism -- available 
within the LVL2 
0.01~s long trigger latency. We recall 
here that the luminosity detector angular acceptance is covered fully by the 
acceptance of the ATLAS silicon tracker \cite{ATLAS},

\item the energies, the shape variables and timing of the 
electromagnetic/hadronic 
clusters in the LAr calorimeter in the ROI-restricted ($\phi, \eta$) regions -- 
available within the LVL2 trigger latency,

\item the reconstructed hits in the LUCID detector \cite{ATLAS} -- available 
within the LVL2 trigger latency,

\item the reconstructed hits in the BCM (Beam Current Monitor)\cite{ATLAS}  --
available within the ~1~s long Event Filter (EF) trigger latency,

\item the precise parameters of the vertex-constrained tracks and the 
multiplicity  of all the charged particles with $p_T > 0.4$~GeV/c produced 
within the range of pseudorapidity $|\eta| <2.5$ -- available within the  EF 
trigger latency.

\end{itemize}
\noindent
It is assumed further that the following host detector performance requirements 
will be fulfilled: 

\begin{itemize}
\item the pion/electron rejection factor of 10 at both the LVL2 and the EF 
trigger levels,

\item full identification of bunch crossings in which the charged particles are 
traversing the LUCID detector ($5.5 \leq |\eta| \leq 6.5$),

\item full identification of bunch crossings  in which the charged particles are 
traversing the BCM ($3.9 \leq |\eta| \leq 4.1$) detector.
\end{itemize}

In the studies presented below the host detector measurement and event selection 
biases are not taken into account. 
The LVL2 and the EF event selection efficiencies, the reconstruction and the 
electron/positron identification efficiencies, for the specified above values of 
the electron/pion rejection factors, are thus set to be equal 1. It is further 
assumed that the lepton pair momentum vectors are reconstructed with infinite 
precision. 

Detailed studies based on realistic host detector performance simulations,  
indispensable if the proposed method is endorsed by the host detector 
collaboration, are both external to the scope of this paper and, more 
importantly, of secondary importance. The luminosity measurement method 
presented in this series of papers was designed such that all the host detector 
measurement biases, resolution functions and efficiencies can be determined 
directly using the data coming from  the host detector recorded event samples.  
 
The selection algorithms are discussed below for the two settings of the 
strength of a uniform solenoidal magnetic field: $B~=~0$ and $B~=~2$~T which are 
labeled respectively as $\bf{B0}$ and $\bf{B2}$. These two magnetic field 
configurations correspond to the zero current and the nominal current of 
the ATLAS central tracker solenoid.

\section{Event Selection Model}
\label{sec:selection}

\subsection{Level 1 Trigger}

The luminosity detector trigger logic analyses the pattern of hits on a
bunch-by-bunch basis. In the first step it selects only the ``low multiplicity 
bunch crossing" candidates \cite{second} i.e. the crossings with less than $N_0$ 
hits in both the $z > 0$ and the $z < 0$ sections of the luminosity 
detector\footnote{The $N_0$ parameter depends upon  the processing power of  
FPGA-based electronics.}.

In the second step, for each low multiplicity bunch crossing, the track 
segments are formed and time-stamp validated. The time-stamp validation of the 
track segments consist of assigning  each of them to one of the following two 
classes: the ``in-time segments" and the ``out-of-time segments". These and the 
subsequent LVL1 selection steps are distinct for the $\bf{B0}$ and $\bf{B2}$ 
magnetic field configurations.  

\subsubsection{$\bf{B0}$ Case}
\label{sec:evseltrigB0}

In the $\bf{B0}$ case a track segment is formed by any combination of the hits,  
$i_1$,  $i_2$ and $i_3$, in the three detector planes satisfying the following 
requirements: 
 
$$min(| i_1 - i_3 |, | i_1 - i_3 +3142 |, | i_3 - i_1 +3142 | ) < i_{cut},$$
\centerline{AND}\\
$$\left\{\frac{i_3+i_1}{2}+ i_{cut} > i_2 > \frac{i_3+i_1}{2}-i_{cut} 
{\rm\ \ \}
\underline{{\rm\ if\ } | i_1 - i_3 | < 1571 \ }}\right., 
$$
\centerline{OR}\\
$$ 
\left.\frac{i_3+i_1 + 3142}{2} + i_{cut} < i_2 < \frac{i_3+i_1 +3142}{2}-i_{cut} \right.\\
$$
$$
\left. \underline{{\rm\ if\ } | i_1 -  i_3 | > 1571, ~ i_1 + i_3 < 3142\ }\right., 
$$
{\rm\ \ \   }
$$
\centerline{OR}\\
$$
$$ 
\left.\frac{i_3+i_1}{2} - 1571+ i_{cut} < i_2 < \frac{i_3+i_1}{2} -1571-i_{cut} \right.\\
$$
$$
\left. \underline{{\rm\ if\ } | i_1 -  i_3 | > 1571,~ i_1 + i_3 \geq 3142\ }\right\}.
$$
The $i_{cut}$ parameter value  is driven by  the luminosity detector thickness 
expressed in units of the radiation length.
 
The in-time segments have the time stamps within the time window of $\Delta 
t_{B0}$ width and an off-set of $t0_{B0}$ with respect to the bunch crossing 
time stamp. All the other track segments are assigned  to the out-of-time 
class\footnote{For a detailed discussion of the timing of the luminosity 
detector signals see \cite{second}.}. In the $\bf{B0}$ case the width of the 
time window reflects both the longitudinal LHC bunch size and  the radial size 
of the luminosity detector.

A bunch crossing is selected by the luminosity detector LVL1 trigger as the 
``2+0" candidate if there are exactly two in-time track segments specified in 
terms of the hit triples: $i_{1,1}$, $i_{1,2}$, $i_{1,3}$ ($i_{2,1}$, $i_{2,2}$, 
$i_{2,3}$) in the left (right) detector part and no in-time track segments in 
the opposite right(left) one. The ``coplanar pair" bunch crossing candidates are
those of the ``2+0" ones in which the hit positions in the first and in the 
third plane satisfy the following conditions:
\begin{equation*}
\left\{
\begin{array}{ccl}
(i_{1,1}+i_{1,3}) \leq 3142  &\wedge& (i_{2,1}+i_{2,3}) <
(i_{1,1}+i_{1,3}) + a_0 \\
(i_{1,1}+i_{1,3}) \leq 3142 &\wedge& (i_{2,1}+i_{2,3}) > 
(i_{1,1}+i_{1,3}) + b_0 \\
(i_{1,1}+i_{1,3}) > 3142 &\wedge& (i_{2,1}+i_{2,3}) < 
(i_{1,1}+i_{1,3}) - b_0  \\ 
(i_{1,1}+i_{1,3}) > 3142 &\wedge& (i_{2,1}+i_{2,3}) > 
(i_{1,1}+i_{1,3}) - a_0. 
\end{array}
\right.
\label{eq:eqline}
\end{equation*}
The above conditions represent the algorithmic, LVL1 trigger 
implementation of the coplanar particle pair selection procedure for the 
$\bf{B0}$ magnetic field configuration discussed in \cite{second}. 

A bunch crossing is selected  as the ``silent bunch crossing" candidate if there 
are no in-time track segments in both the left and right side of the 
luminosity detector. 

\subsubsection{$B2$  Case}
\label{sec:evseltrigB2}

In the $\bf{B2}$ case a track segment is formed by any combination of the hits 
$i_1$, $i_2$, $i_3$ in the three detector planes which satisfy the 
requirements outlined below: 

$$min(| i_1 - i_3 |, | i_1 - i_3 +3142 |, | i_3 - i_1 +3142 | ) < i_{cut}+i_{helix},$$
\centerline{AND}\\
$$\left\{\frac{i_3+i_1}{2}+ i_{cut} > i_2 > \frac{i_3+i_1}{2}-i_{cut} 
{\rm\ \ \}
\underline{{\rm\ if\ } | i_1 - i_3 | < 1571 \ }}\right., 
$$
\centerline{OR}\\
$$ 
\left.\frac{i_3+i_1 + 3142}{2} + i_{cut} < i_2 < \frac{i_3+i_1 +3142}{2}-i_{cut} \right.\\
$$
$$
\left. \underline{{\rm\ if\ } | i_1 -  i_3 | > 1571, ~ i_1 + i_3 < 3142\ }\right., 
$$
{\rm\ \ \   }
$$
\centerline{OR}\\
$$
$$ 
\left.\frac{i_3+i_1}{2} - 1571+ i_{cut} < i_2 < \frac{i_3+i_1}{2} -1571-i_{cut} \right.\\
$$
$$
\left. \underline{{\rm\ if\ } | i_1 -  i_3 | > 1571,~ i_1 + i_3 \geq 3142\ }\right\}.
$$
The magnetic field strength dependent $i_{helix }$ value drives the effective 
momentum cut-off of the charged particles reaching the luminosity detector. The 
in-time track segments have the time-stamps  within the time window of $\Delta 
t_{B2}$ width and the off-set of $t0_{B2}$~ns with respect to the bunch crossing 
time-stamp. The timing parameters are different for the $\bf{B0}$ and $\bf{B2}$
cases because for a given charged particle its track length measured from 
the interaction vertex to the entry point into the fiducial volume of the 
luminosity detector is different in these two cases.  

For the equidistant spacing of the $z$-planes the track segment finding and the 
time-stamp validation algorithms remain invariant with respect to the change of 
the magnetic field configuration - its influence is restricted only to the 
parameters of the algorithms. 

As in the $\bf{B0}$ case, a bunch crossing is selected, by the luminosity 
detector LVL1 trigger, as the ``2+0" candidate if there are exactly two in-time 
track segments in the left (right) detector part and no in-time track segments
in the opposite one. However, in the $\bf{B2}$ case a supplementary condition
is required to retain only the opposite charge particle tracks: 
$$( i_{1,1} - i_{3,1} \geq 0  ~~if~~   i_{1,2} - i_{3,2}  < 0)  ~~or~~  
(i_{1,1} - i_{3,1}  < 0  ~~if~~   i_{1,2} - i_{3,2}  \geq 0).$$ 

The ``coplanar pair" bunch crossing candidates are those of the ``2+0" ones in 
which the hits $i_{1,1}$,  $i_{1,3}$ $(i_{2,1}, i_{2,3})$ satisfy the following 
conditions:
\begin{equation*}
\left\{
\begin{array}{ccl}
| i_{1,1} - i_{2,1}| &<& a_2\cdot(  i_{1,1} - i_{1,3} +  i_{2,1} - i_{2,3} )\\ 
| i_{1,1} - i_{2,1}| &<& b_2\cdot( i_{1,1} - i_{1,3} +  i_{2,1} - i_{2,3})+e_2 \\
| i_{1,1} - i_{2,1}| &>& c_2\cdot( i_{1,1} - i_{1,3} +  i_{2,1} - i_{2,3}) \\
| i_{1,1} - i_{2,1}| &>& d_2\cdot( i_{1,1} - i_{1,3} +  i_{2,1} - i_{2,3})-f_2. 
\end{array}
\right.
\label{eq:diamond}
\end{equation*}
The above conditions define the LVL1 trigger algorithm for the coplanar particle 
pair selection procedure for the $\bf{B2}$ magnetic field configuration 
(see \cite{second} for detailed discussion). 

Similarly to the $\bf{B0}$ case, a bunch crossing is selected as the silent 
bunch crossing if there are no in-time track segments in both parts of the 
luminosity detector. 

\subsection{LVL1 Trigger Bits} 

It is assumed that the following LVL1 trigger bits are sent by the luminosity
detector to the Central Trigger Processor (CTP) of the host detector:
\begin{itemize}
\item the ``coplanar pair candidate" (CPC) trigger bit, 
\item the ``2+0" (two plus zero - TPZ) trigger bit,   
\item the ``silent bunch crossing" (SBC) trigger bit,
\item the ``low multiplicity bunch crossing" (LMBC) trigger bit. 
\end{itemize}
They are assumed to be broadcasted on the bunch-by-bunch  basis. For the CPC 
(TPZ)-trigger-selected bunch crossings  the information on the $\phi$-sector 
positions of the two in-time track segments is delivered the host detector Level 
2 trigger algorithms using  the ROI  mechanism. The CPC trigger is not prescaled 
at the CTP level. 

While the CPC-selected bunch-crossing data are used in the luminosity 
determination, the CTP-prescaled TPZ, SBC and LMBC ones, together with the 
CTP-selected random bunch crossings data, are  used for a precision monitoring 
of the luminosity detector performance. In particular, in all the aspects which 
require correlating of the luminosity detector signals with the host detector 
ones on bunch-by-bunch basis.  

\subsection{Level 2 Trigger}

The next step in the event selection chain is based entirely on the host 
detector LVL2 trigger algorithms. In the following we shall discuss only the 
LVL2 selection criteria for the LVL1-accepted coplanar pair candidate events,  
for which the CPC trigger bit was set to one\footnote{The LVL2 and EF processing 
of the monitoring events will be discussed later while addressing the precision 
of the proposed luminosity measurement method.}. As before, the selection 
criteria are different for each of the two magnetic field configurations. 

\subsubsection{$\bf{B0}$ Case}

In the first step of the LVL2 selection algorithm chain a search for a narrow 
energy cluster of the total energy above 1~GeV is performed. The search is 
confined to the electromagnetic calorimeter $\phi$-sectors pointed out by the 
luminosity detector LVL1 ROI. If two electromagnetic clusters are found and if 
their timing, determined from the pulse-shapes in the channels belonging to 
clusters, is compatible with the track segment time-stamp the event is retained 
for the subsequent selection steps. The next step links the luminosity detector 
track segments to the  Silicon Tracker (SCT) hits. If the linking is successful 
and if the corresponding SCT track segments cross each other in a space point 
within the proton bunch overlap IP region then the reduced $vertex$ acoplanarity 
$\delta\phi_r$ is recalculated using the SCT hits. Events are retained for 
further processing if $\delta\phi_r  < \delta\phi_r^{cut}$ is fulfilled. The 
subsequent algorithm verifies if both clusters pass the electron selection 
criteria by analysing their  lateral and longitudinal shape. Events passing all 
the above selection steps will be called in the following the LVL2 trigger 
``inclusive electron-positron pair" candidates. In the subsequent LVL2 step the 
LUCID detector signals are analysed and an event is selected as the LVL2 
``exclusive electron-positron  pair" candidate if there are no in-time particle 
hits in the LUCID tubes\footnote{The LUCID ``particle hit" is expected to be set 
at a sufficiently high discriminator threshold to be as much as possible 
noise-free.}.

\subsubsection{$\bf{B2}$ Case}

The only difference in the selection steps of the LVL2 exclusive 
electron-positron pair candidates for the $\bf{B2}$ case is that the cluster 
energy cut is no longer imposed. This is because the equivalent cut, was already 
made by the  LVL1 trigger. 

\subsection{Event Filter}

The LVL2 selection of the exclusive electron-positron pair candidate events is 
subsequently  sharpened at the Event Filter (EF) level. Similarly to the LVL1 
and LVL2 cases the selection criteria for both configurations of the magnetic 
field slightly differ. 

\subsubsection{$\bf{B0}$ case}

In the first step of the EF selection algorithm chain events with any 
reconstructed particle tracks within the tracker fiducial volume other than the 
tracks of the electron pair candidate are rejected. Subsequently, the reduced 
{\it vertex} acoplanarity $\delta\phi_r$ is recalculated using the re-fitted 
vertex constrained values of the parameters of the lepton tracks and the 
$\delta\phi_r < \delta\phi_r^{cut}$ cut is sharpened. Next, the EF electron 
selection algorithms exploiting full information coming from both the tracker 
and the LAr calorimeter are run and a further rejection of hadrons mimicking the 
electron signatures is performed. Finally, the event exclusivity requirement is 
sharpened by demanding no particle hits in the BCM in a tight time window.  

\subsubsection{$\bf{B2}$ Case}

There are only two differences in the selection steps of the exclusive lepton 
pair candidates in the $\bf{B2}$ field configuration case. These are: 
\begin{itemize}
\item a replacement of the exclusivity criterion using the reconstructed track 
segments by the corresponding one based on the reconstructed tracks with the 
transverse momentum  $p_{T} >  0.4$~GeV/c,
\item a restriction of the transverse momentum of a pair to the region  
$p_{T,pair}~<~0.05$~GeV/c.  
\end{itemize}

\section{Optimisation of the LVL1 Trigger} 

\subsection{Algorithm parameters} 

The LVL1 trigger  optimisation goal is to determine an optimal set of the 
algorithm parameters. These parameters, defined in Section \ref{sec:evseltrigB0} and \ref{sec:evseltrigB2},  specify:
\begin{itemize}
\item the definition of the track segments, 
\item the classification of the track segments into the in-time and out-of-time
classes, 
\item the coplanarity of particle pairs (at the interaction vertex). 
\end{itemize}

An optimal set of parameters maximises the rate of the exclusive coplanar pair 
candidate events while retaining the overall luminosity detector LVL1 trigger 
rate at ${\cal O}(1)$~kHz level. The latter restriction takes into account the 
present capacity of the ATLAS TDAQ system and assumes that at most $\sim 2\%$ of 
its throughput capacity can be attributed to the luminosity detector triggered 
events. For these events the event record length and the LVL2 and the EF filter 
processing times are significantly smaller than for any other physics triggers 
of the present host detector LVL1 trigger menu. Therefore, the strain on the 
LVL2 and EF throughputs is  negligible. 

The LVL1 trigger algorithm parameters were optimized by simulating the selection 
process for large samples of the bunch crossings containing the signal and the 
background events. The LVL1 trigger algorithms assigned  the 0 or 1 values to 
the CPC, SBC and LMBC LVL1 trigger bits for every bunch crossing\footnote{The 
electron-positron pair signal events were generated with the LPAIR \cite{LPAIR} 
generator. This generator was upgraded to suit our needs (see \cite{first} for 
details). For the simulations of the minimum bias events the PYTHIA 
\cite{PYTHIA} event generator was used. As discussed in details in our earlier 
papers \cite{first}, \cite{second}, the studies were based on simplified methods 
of particle tracking in the detector magnetic field, parametrised simulation of 
their multiple scattering in the dead material, and on conservative estimation 
of the effects of the photon radiation by electrons.}.

The following sets of parameters maximises the signal to background ration while  
retaining the overall luminosity detector LVL1 trigger rate below 2~kHz level: 
\begin{itemize}
\item track segment definition:  
$$
i_{cut}=20, i_{helix }= 130,
$$
\item selection of in-time track segments:  
$$
 \Delta t_{B0}= 1.5~ns, 
\Delta t_{B2} = 4~ns, t0_{B0}  = t0_{B2} =  19~ns,
$$
\item pair acoplanarity selection: 
$$
a_0 =  3202, b_0 = 3082,
$$
$$
a_2 = 5.97, b_2 = 1.78, c_2 = 4.97,$$ 
$$
d_2 = 30.97, e_2 = 953.4, f_2 = 6630.0.
$$
\end{itemize}

\begin{figure}[ht]
\begin{center}
\setlength{\unitlength}{1mm}
\begin{picture}(130,120)
\put(0,65){\makebox(0,0)[lb]{
\epsfig{file=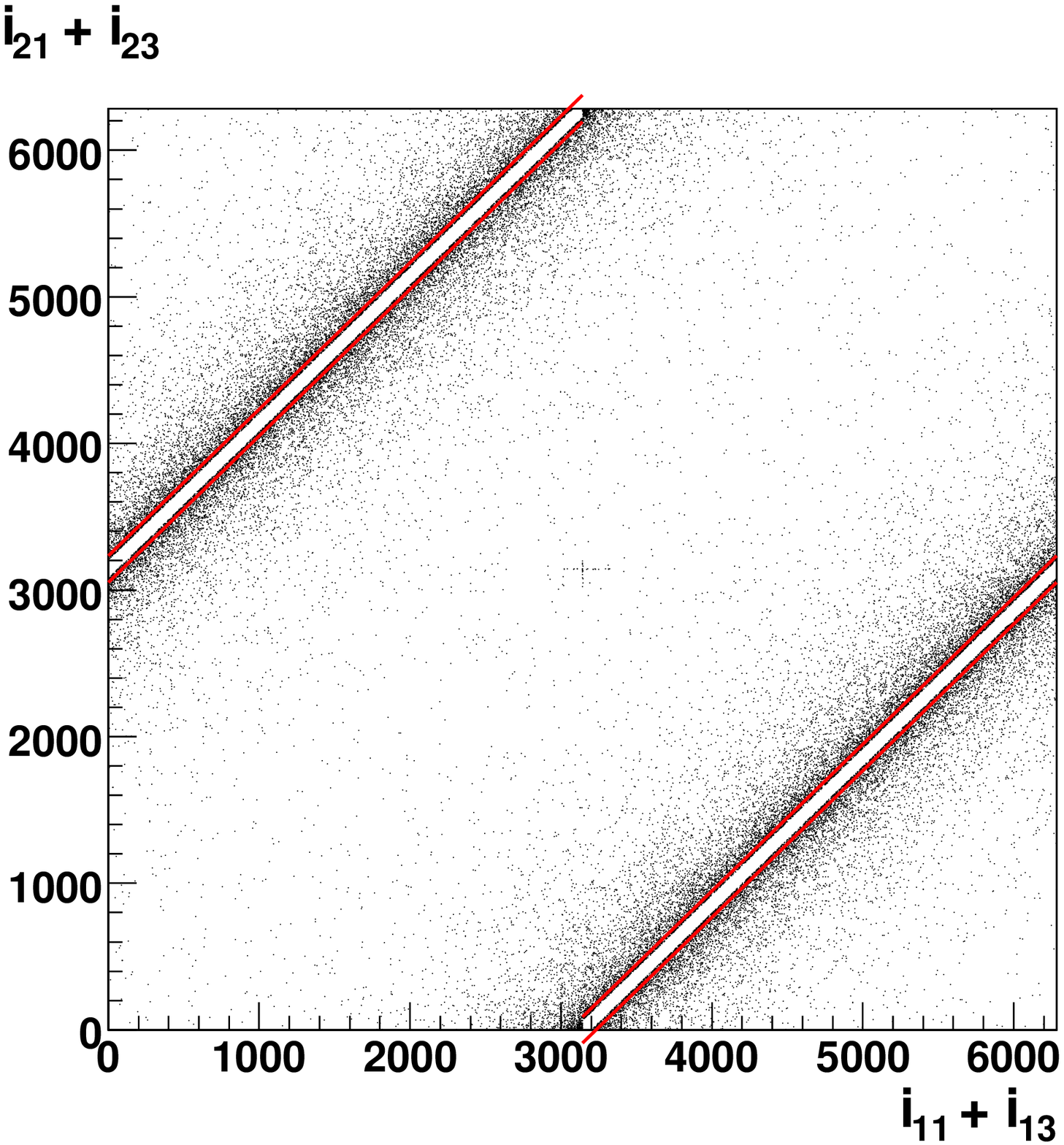, width=60mm,height=60mm}
}}
\put(0, 3){\makebox(0,0)[lb]{
\epsfig{file=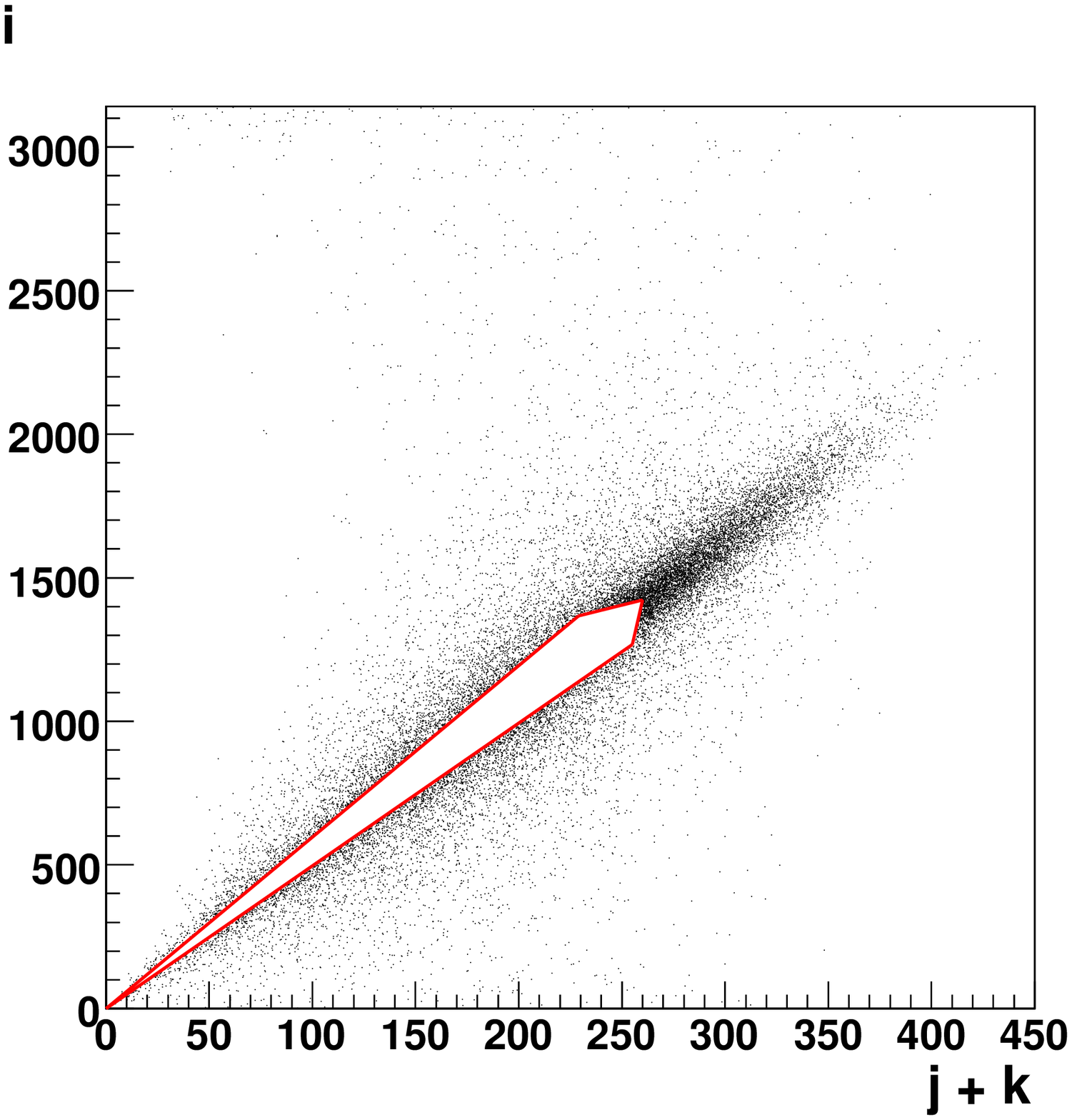, width=60mm,height=60mm}
}}
\put(65,65){\makebox(0,0)[lb]{
\epsfig{file=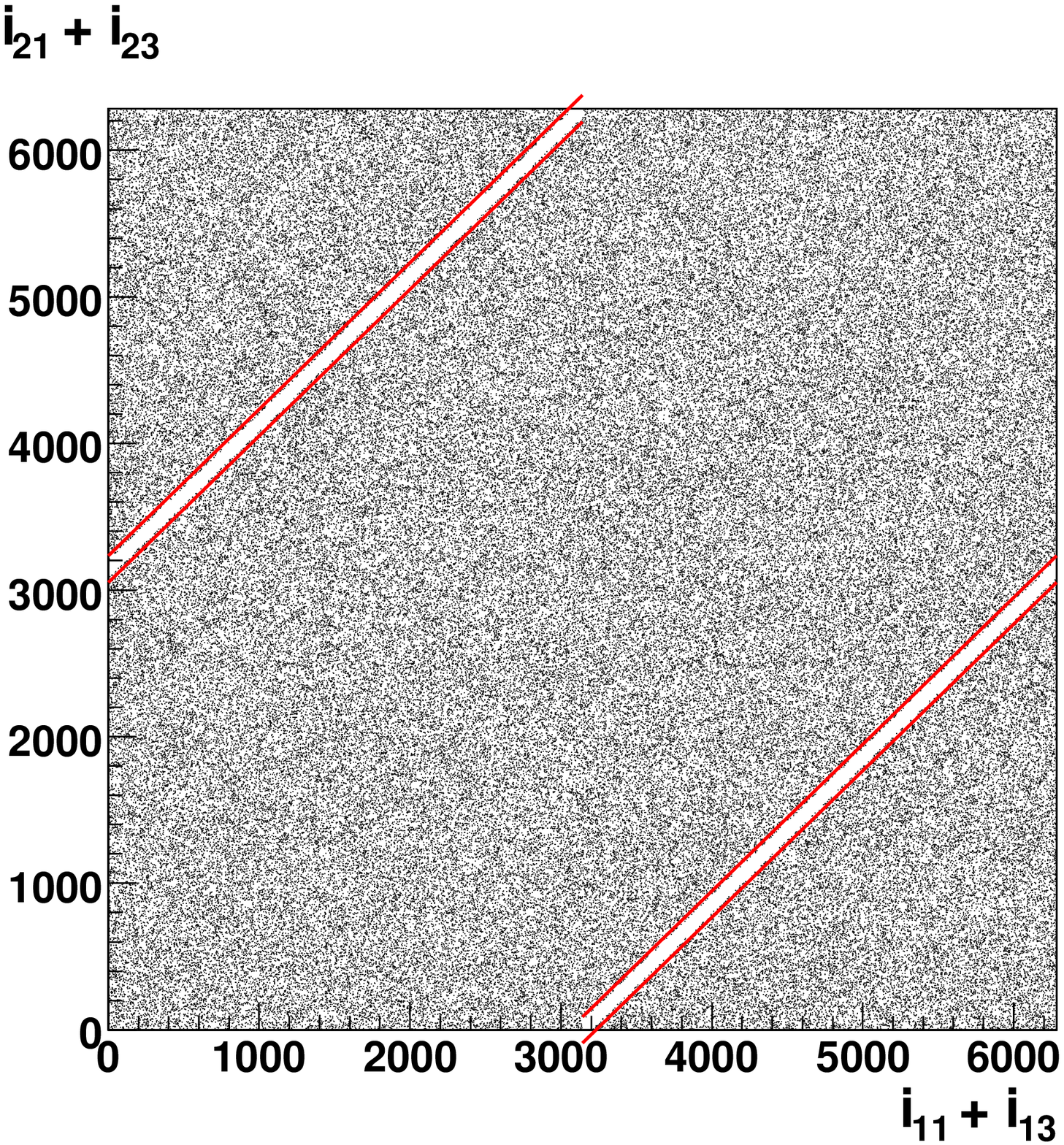, width=60mm,height=60mm}
}}
\put(65, 3){\makebox(0,0)[lb]{
\epsfig{file=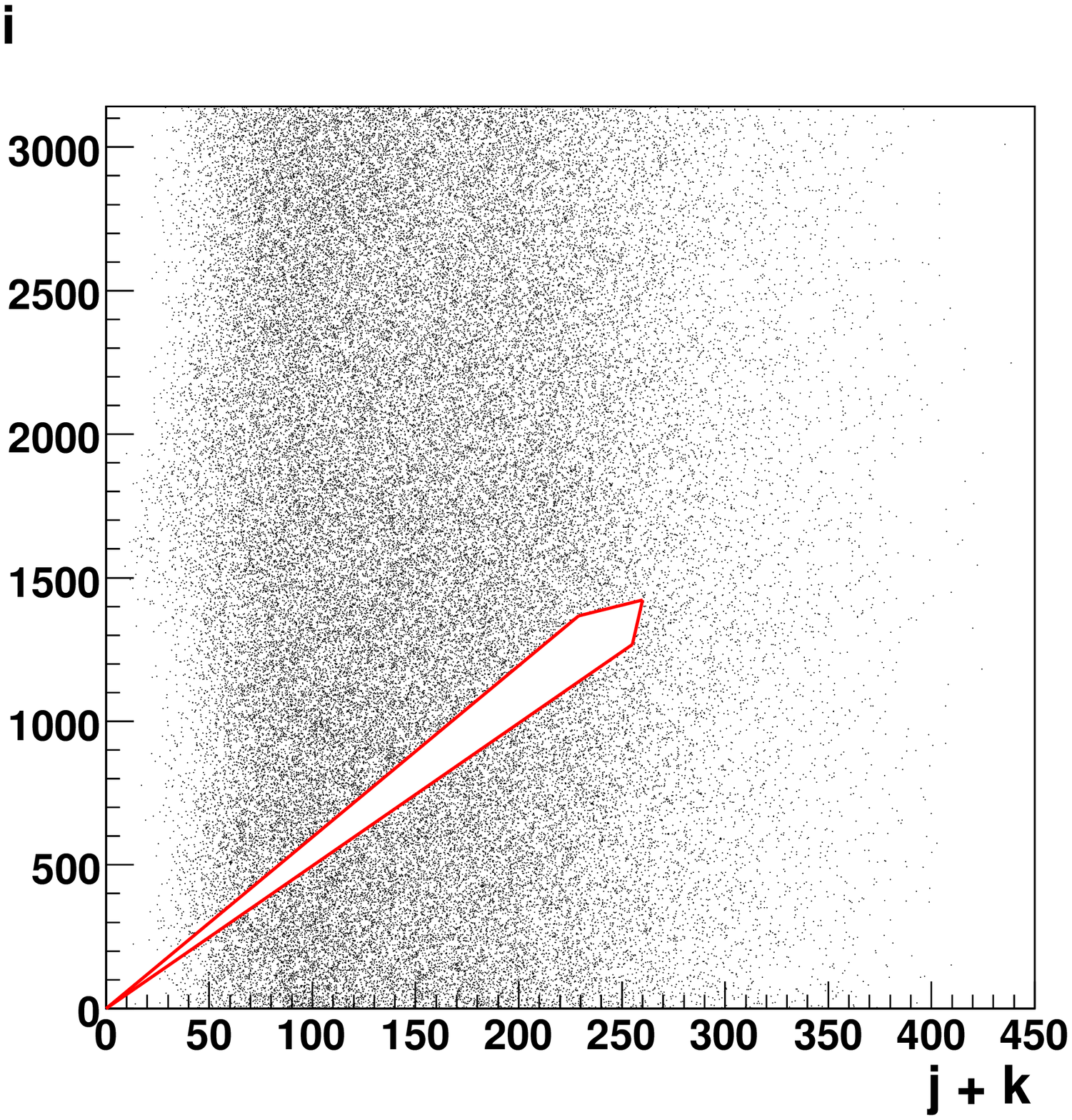, width=60mm,height=60mm}
}}
\put(95,62){\makebox(0,0)[cb]{\bf (b)}}
\put( 33,62){\makebox(0,0)[cb]{\bf (a)}}
\put( 33,-2){\makebox(0,0)[cb]{\bf (c)}}
\put(95,-2){\makebox(0,0)[cb]{\bf (d)}}

\put(110,121){\makebox(0,0)[cb]{\bf {\small B = 0~T}}}
\put( 46,121){\makebox(0,0)[cb]{\bf {\small B = 0~T}}}
\put( 46,59){\makebox(0,0)[cb]{\bf {\small B = 2~T}}}
\put(110,59){\makebox(0,0)[cb]{\bf {\small B = 2~T}}}
\end{picture}
\end{center}
\caption{The LVL1 trigger selection of coplanar electron-positron pair CPC-events: 
-- (a) the correlation of the sum of the indices of 
the hit positions of the first particle in the first and the third plane: 
$(i_{1,1}+i_{1,3})$ with the corresponding sum $(i_{2,1}+i_{2,3})$ for the 
second particle,  for the LPAIR signal events and for the $\bf{B0}$ field 
configuration; -- (b) as (a) but for the PYTHIA background events,          
-- (c) the correlation between: (1) $j+k$, where $j = i_{1,1} - i_{1,3}$ and 
$k = i_{2,1} - i_{2,3}$ are the differences of the hit positions in the first 
and in the third plane left by the first and by the second particle, 
respectively, and (2) the difference between the hit positions in the first  plane 
for the first and second particle, $i = i_{1,1} - i_{2,1}$, for the LPAIR  
signal events;  -- (d) as (c) but for the PYTHIA background events. 
For events inside the region marked by the lines the ``coplanar pair 
candidate",  (CPC) trigger bit is set to one and to zero elsewhere.}
\label{plot15}
\end{figure}

The $i_{helix }= 130$ corresponds to the effective low-momentum cut-off for
particles producing a track segment in the luminosity detector of ~1~GeV/c. The
$i_{cut}=20$ reflects the assumed thickness of the detector planes of 0.1$X_0$ 
each. The increase of the width of the in-time window, for the $\bf{B2}$ case 
reflects the dispersion of the helix length of the charged particle trajectories 
between the interaction vertex and the first plane of the luminosity detector. 
The optimal selection region of coplanar pairs is illustrated in Fig. 
\ref{plot15} for a chosen set of selection parameters. 

%
\clearpage
Note, that for the $\bf{B2}$ configuration a sizable fraction of the 
electron-positron pair signal events is rejected by the track segment validation 
criteria. In these events at least one of particles has the momentum lower than 
$\sim$1~GeV/c. 

The selection efficiency of the coplanar particle pairs is determined using the
reduced acoplanarity, $\delta\phi_r$, variable which is defined as 
$$\delta\phi_r = \delta\phi/\pi,$$
with
$$\delta\phi=\pi-min(2\pi-|\phi_{1}-\phi_{2}|,|\phi_{1}-\phi_{2}|),$$  
where $\phi_{1}$, $\phi_{2}$ are the azimuthal angles of the particles at the 
interaction vertex.

\subsection{Algorithm efficiency}

\begin{figure}[ht]
\begin{center}
\setlength{\unitlength}{1mm}
\begin{picture}(130,150)
\put(0,65){\makebox(0,0)[lb]{
\epsfig{file=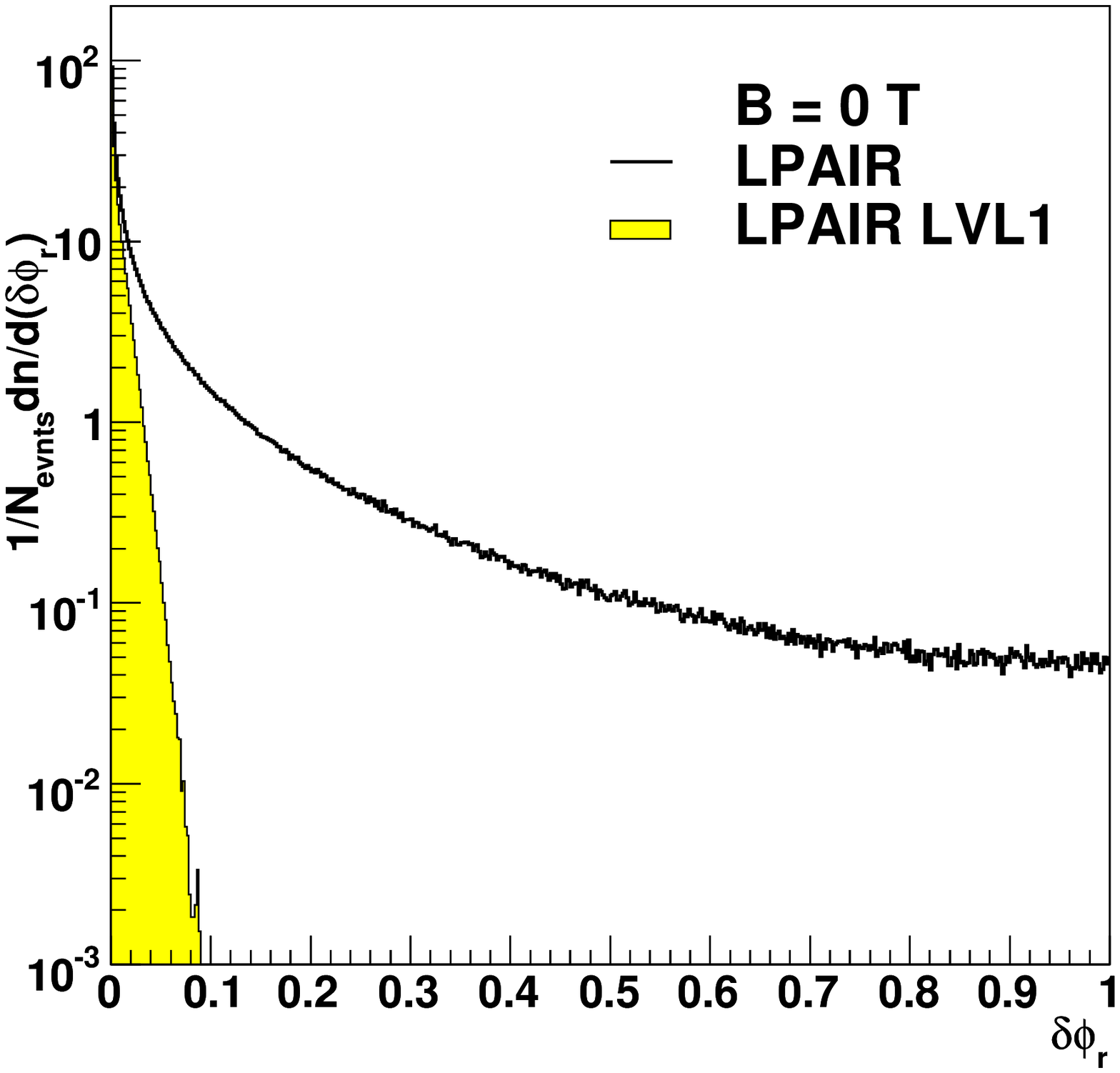, width=65mm,height=65mm}
}}
\put(0, 0){\makebox(0,0)[lb]{
\epsfig{file=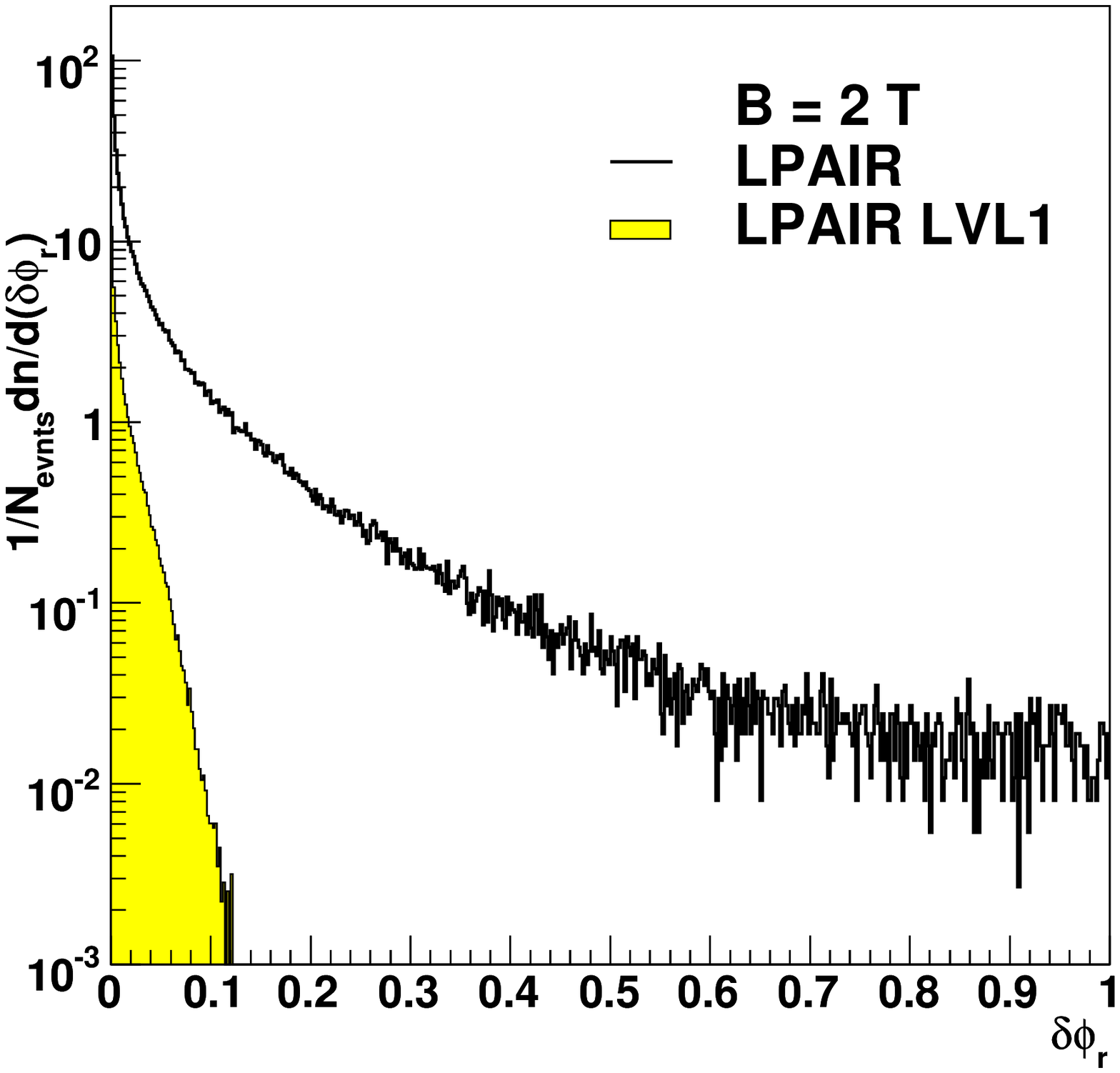, width=65mm,height=65mm}
}}
\put(65,65){\makebox(0,0)[lb]{
\epsfig{file=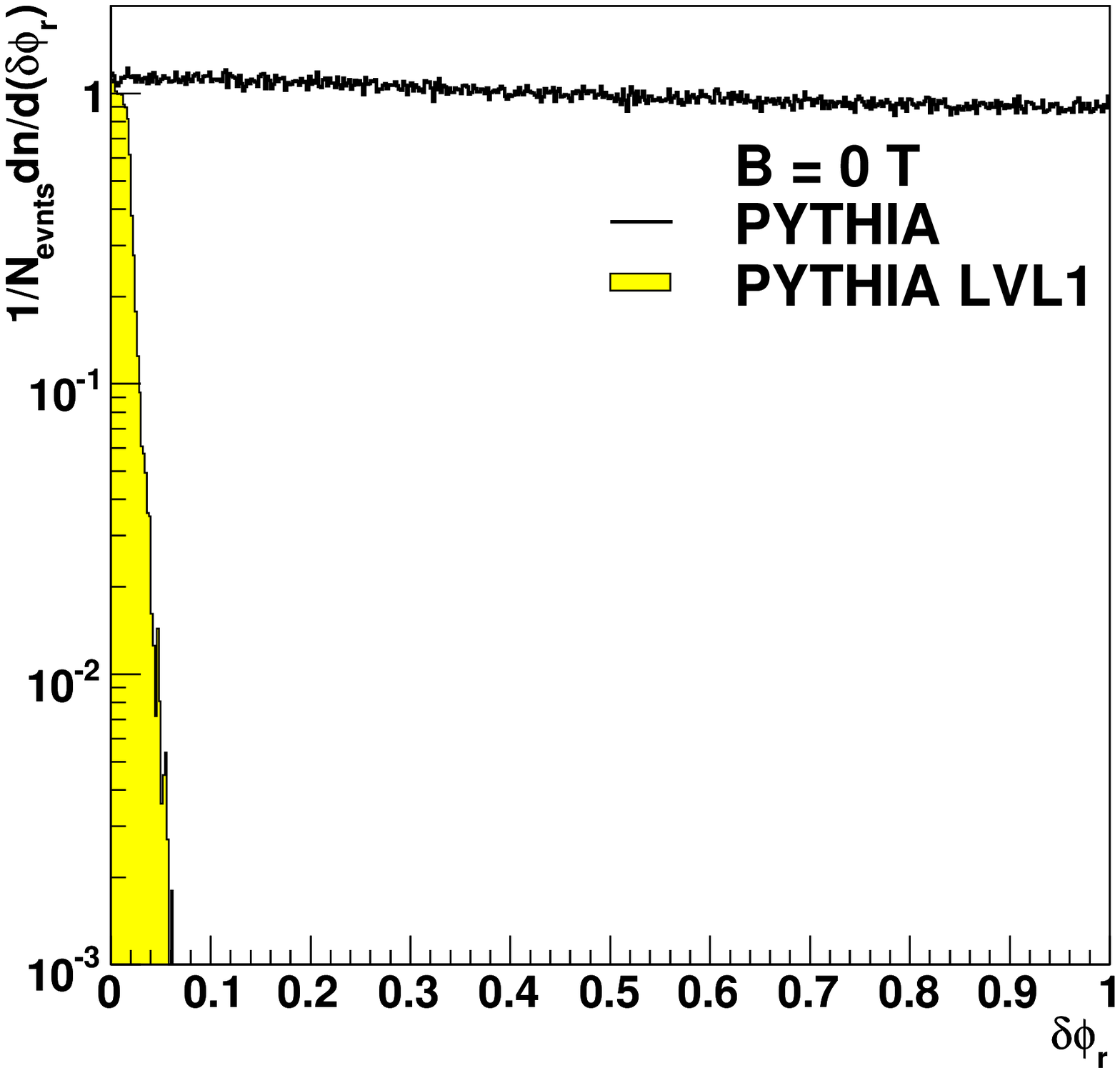, width=65mm,height=65mm}
}}
\put(65, 0){\makebox(0,0)[lb]{
\epsfig{file=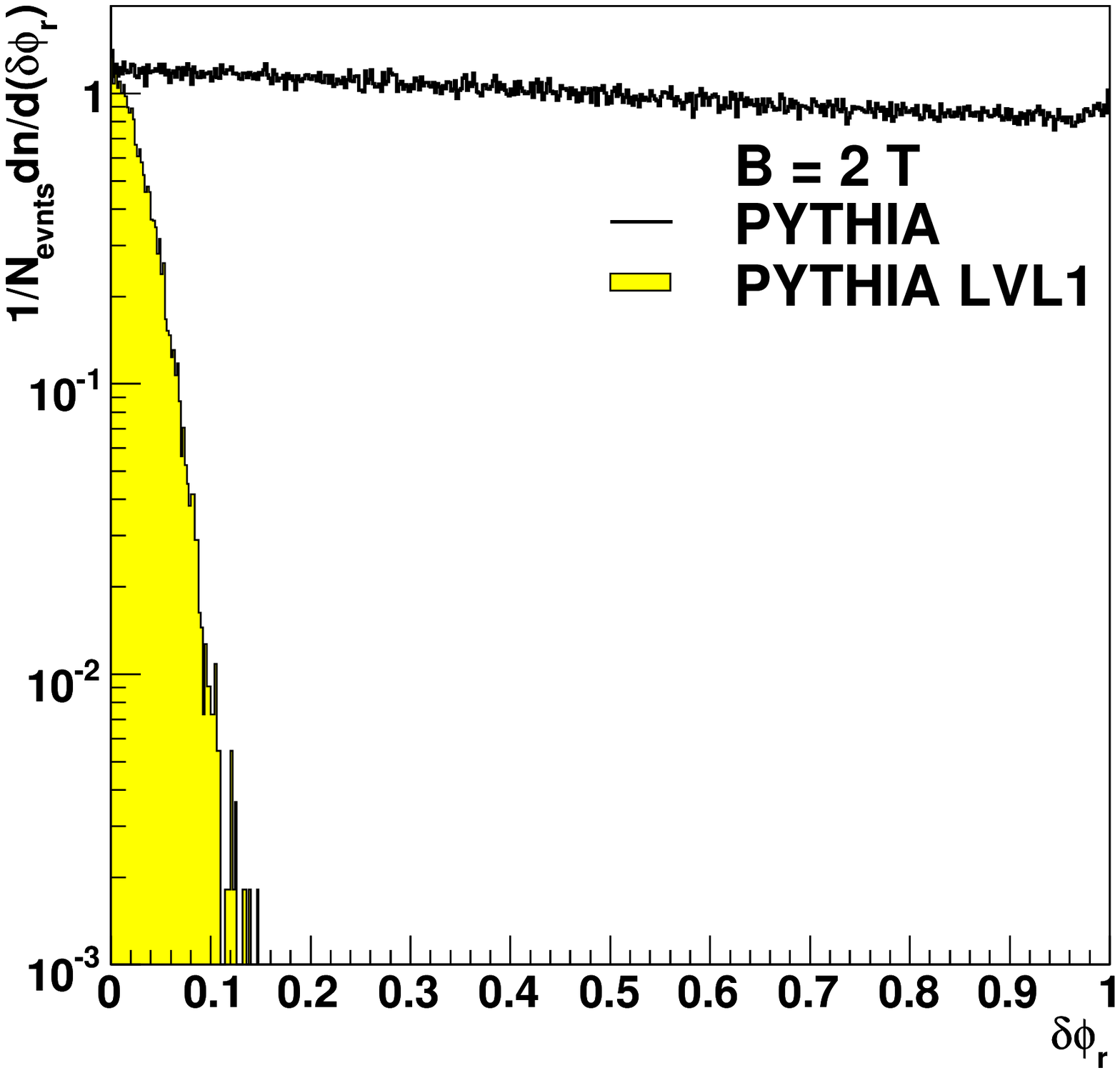, width=65mm,height=65mm}
}}
\put( 33,63){\makebox(0,0)[cb]{\bf (a)}}
\put(95,63){\makebox(0,0)[cb]{\bf (b)}}
\put( 33,-3){\makebox(0,0)[cb]{\bf (c)}}
\put(95,-3){\makebox(0,0)[cb]{\bf (d)}}
\end{picture}
\end{center}
\caption{The study of the LVL1 trigger efficiencies for the samples of the LPAIR 
signal and of the ``2+0" PYTHIA background events for the two magnetic  field 
configurations. The thick solid histogram marks the input reduced acoplanarity 
distributions. The shaded are marks distributions for the LVL1 (CPC-trigger)  accepted events:
-- (a) and (b)  for the $\bf{B0}$ field configuration,
-- (c) and (d)  for the $\bf{B2}$ field configuration. }
\label{plot16}
\end{figure}

The efficiency of the LVL1 trigger algorithms of selecting the coplanar particle 
pairs using the above sets of parameters is illustrated in Fig. \ref{plot16}. 
In this figure the reduced acoplanarity distributions are plotted for the 
signal and background events, the two magnetic field configurations and for the 
initial and  the CPC-trigger selected samples events. In the $\bf{B0}$ case the 
LVL1 algorithms rejects majority of the background events while retaining the 
signal events. In the case of the $\bf{B2}$ field configuration a sizable 
reduction of the electron-positron pair selection efficiency is not related to 
the luminosity detector performance.  It is almost entirely driven by the 
constraint on the luminosity detector position within the host detector fiducial 
volume giving rise to a large probability of  bremsstrahlung of the hard photon 
by the electron or positron on the path between the collision vertex and the 
luminosity detector entry point. 

The efficiencies presented above correspond to the most pessimistic assumptions 
on the luminosity detector performance and on the dead material budget in front 
of the luminosity detector. Firstly, the base-line luminosity detector model was 
used. This model does not employ a precise relative timing in each of the three 
detector planes. Therefore, the event-by-event LVL1-trigger reconstruction of 
the position of the collision vertex, an option discussed in our previous paper 
\cite{second}, was not made. Such a function would significantly improve the 
sharpness of the LVL1 acoplanarity algorithm. Secondly, the studies were made 
for the 0.9$X0$ of dead material in front of the luminosity detector giving a 
rise to large multiple scattering effects. Thirdly, and most importantly, the 
hard photon radiation was assumed to take place at the collision vertex, 
inducing significant loss of efficiency for the $\bf{B2}$ case. Therefore, the 
results in this section represents the most conservative estimate of the 
electron-positron pair event selection efficiency.

\section{Rates}

\begin{figure}[h]
\begin{center}
\setlength{\unitlength}{1mm}
\begin{picture}(130,70)
\put(0,0){\makebox(0,0)[lb]{
\epsfig{file=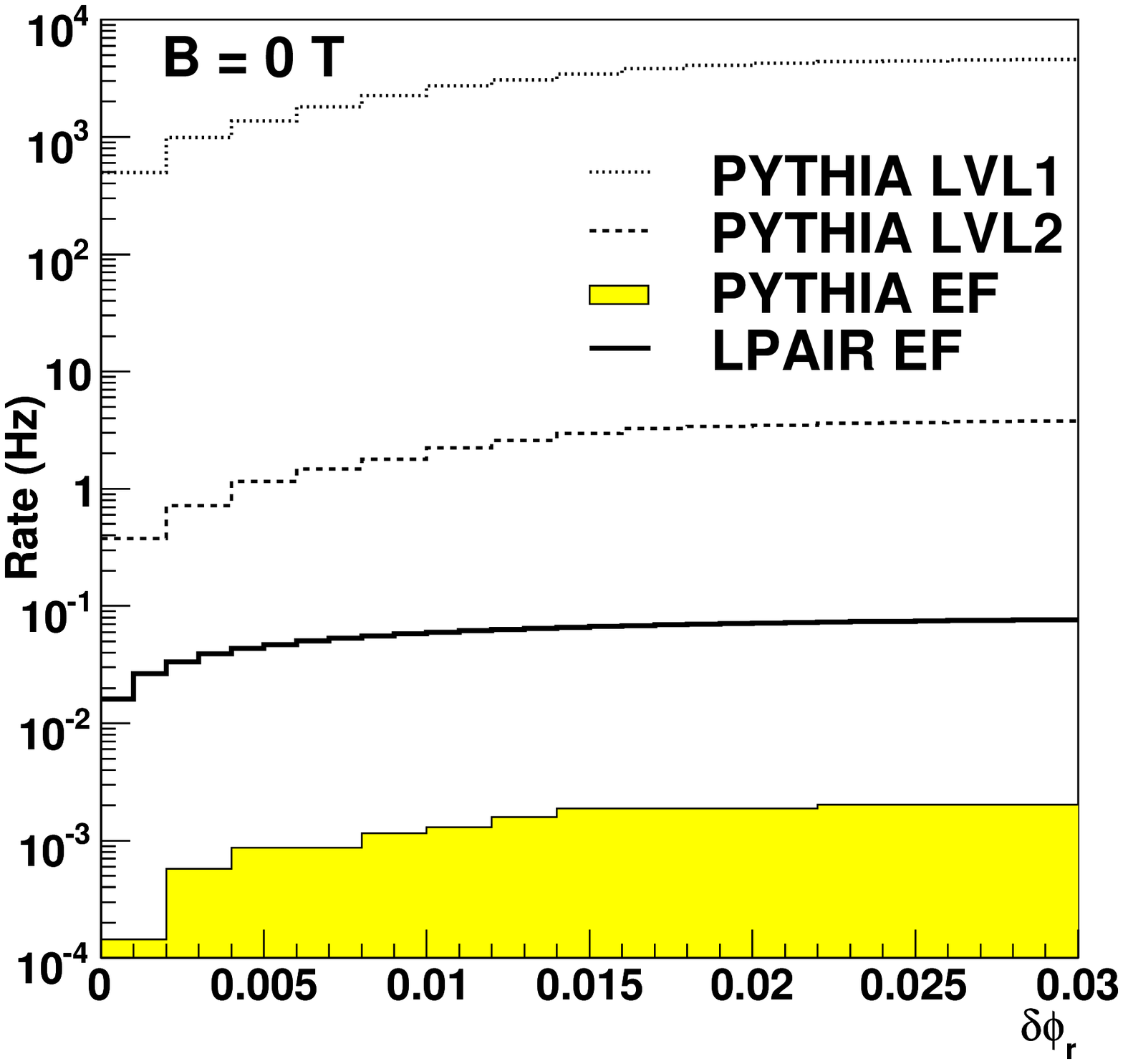, width=65mm,height=65mm}
}}
\put(65,0){\makebox(0,0)[lb]{
\epsfig{file=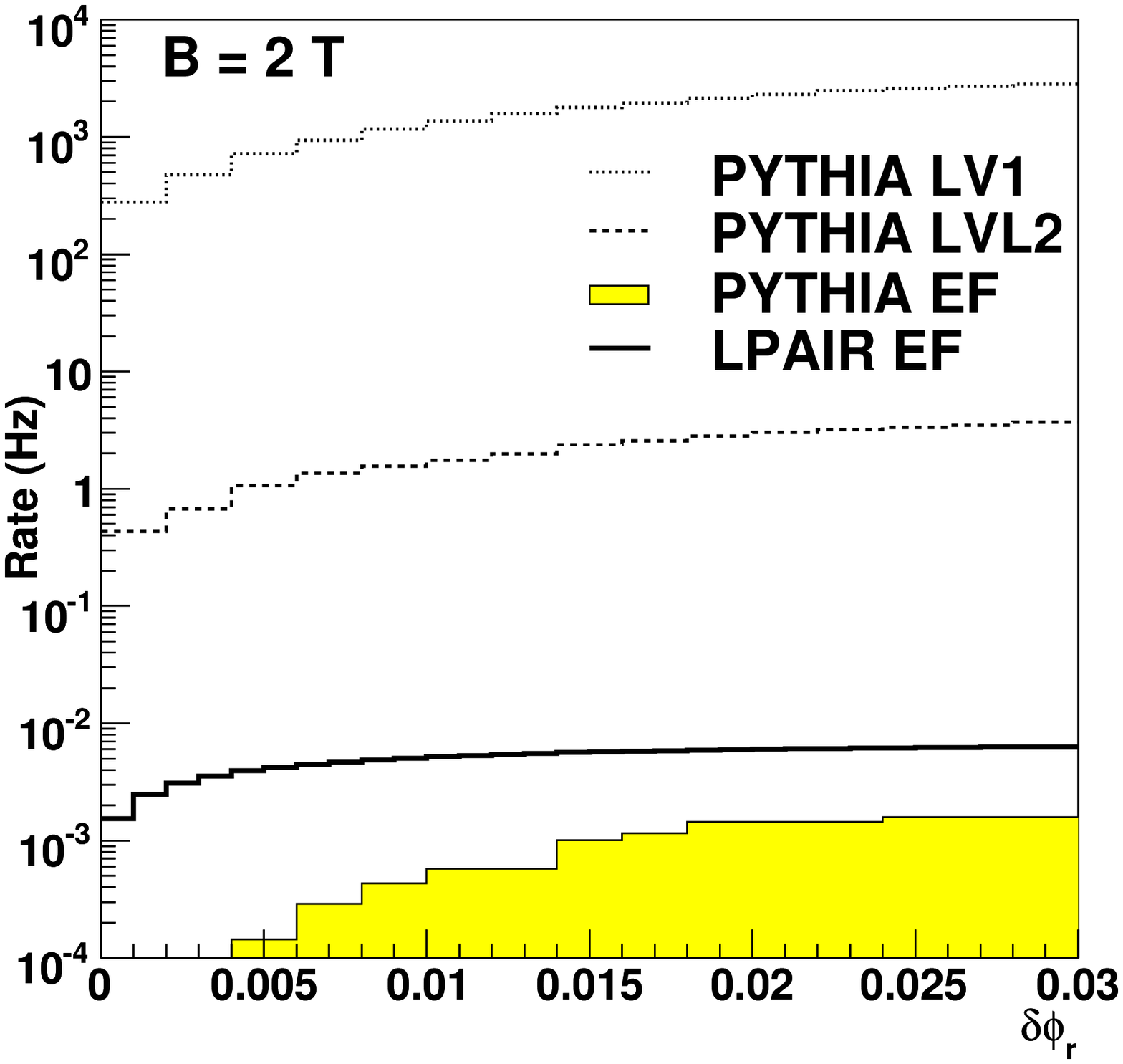, width=65mm,height=65mm}
}}
\put( 33,-5){\makebox(0,0)[cb]{\bf (a)}}
\put(95,-5){\makebox(0,0)[cb]{\bf (b)}}
\end{picture}
\end{center}
\caption{The integrated rate of the background (PYTHIA) events as a function of 
the upper limit on the pair $\delta\phi_r$ at the three consecutive stages
of the event selection process: 
the dotted line -- following the LVL1 decision, the dashed line -- following the 
LVL2 decision, the shaded area - following the EF decision. The integrated rate 
of the signal (LPAIR) events accepted at the EF level is marked with the thick 
solid line. The plots were made for the machine luminosity of $L= 10^{33} s^{-1}cm^{-2}$.
--~(a)~for the $\bf{B0}$ field configuration.
--~(b)~for the $\bf{B2}$ field configuration. }
\label{plot17}
\end{figure}

In Figure \ref{plot17} the integrated rates of the LVL1-accepted coplanar pair 
candidate events are plotted as a function of the upper limit on the pair 
$\delta\phi_r$ for the two magnetic field configurations, for the signal and the 
background events. The distributions for the background events are shown at all 
the stages of the event selection procedure. This plot was made for the machine 
luminosity of $L~=~10^{33}$~cm$^{-2}$~s$^{-1}$ distributed uniformly over all 
the available bunch crossing slots. 

\clearpage

This figure demonstrates that the proton-proton collisions producing coplanar 
electron-positron pairs can be efficiently selected from the background of 
minimum bias events. A reduction of the signal rate seen for the $\bf{B2}$ 
field configuration as compared to the $\bf{B0}$ configuration is driven mainly 
by radiation of hard photons by the electrons/positrons passing the host 
detector dead material. The signal to noise ratio is the largest for the 
smallest reduced acoplanarity cutoff. It underlines the need for a fine 
$\phi$-segmentation of the luminosity detector.  

\begin{figure}
\begin{center}
\setlength{\unitlength}{1mm}
\begin{picture}(130,60)
\put(0,0){\makebox(0,0)[lb]{
\epsfig{file=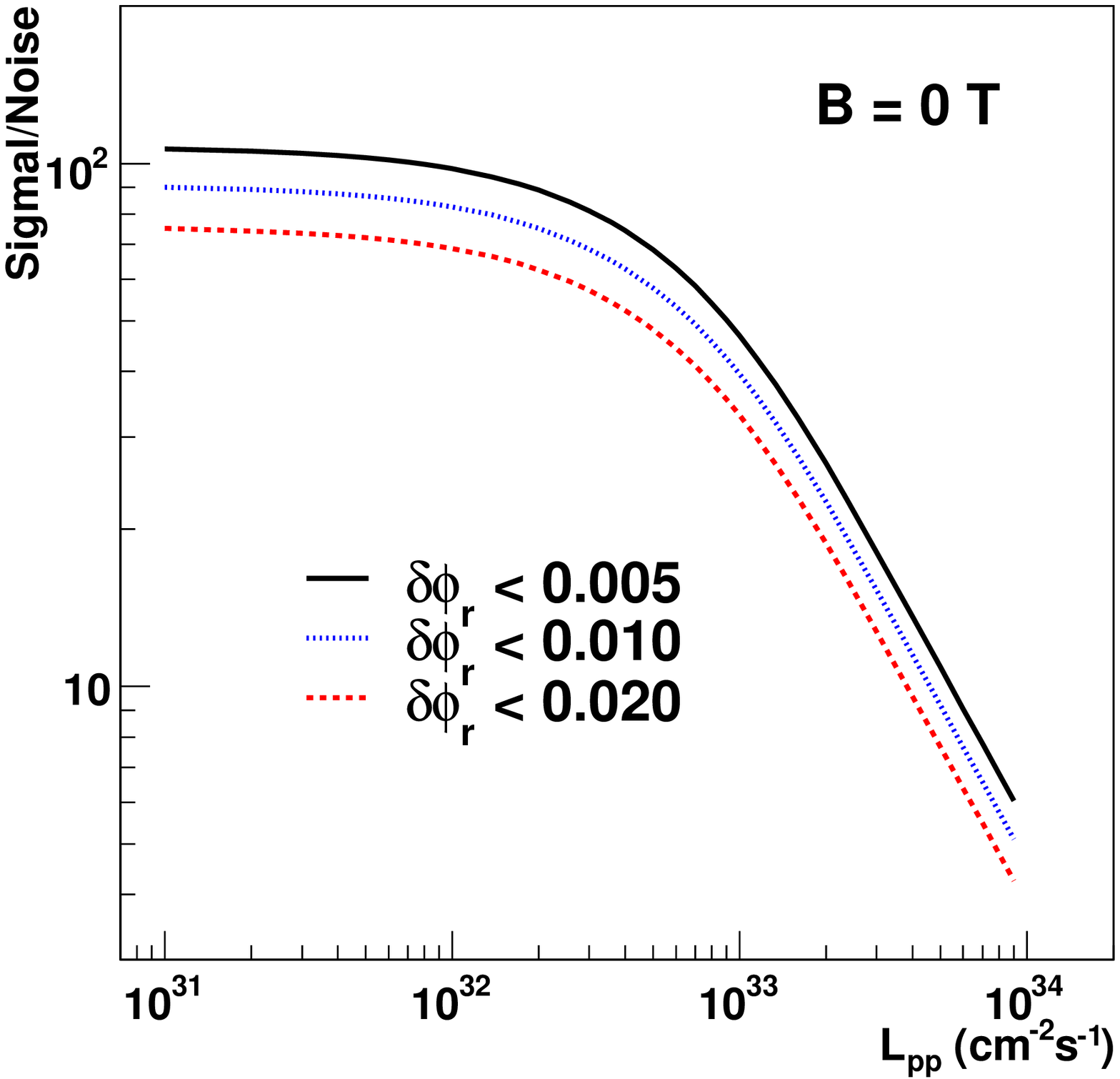, width=65mm,height=65mm}
}}
\put(65,0){\makebox(0,0)[lb]{
\epsfig{file=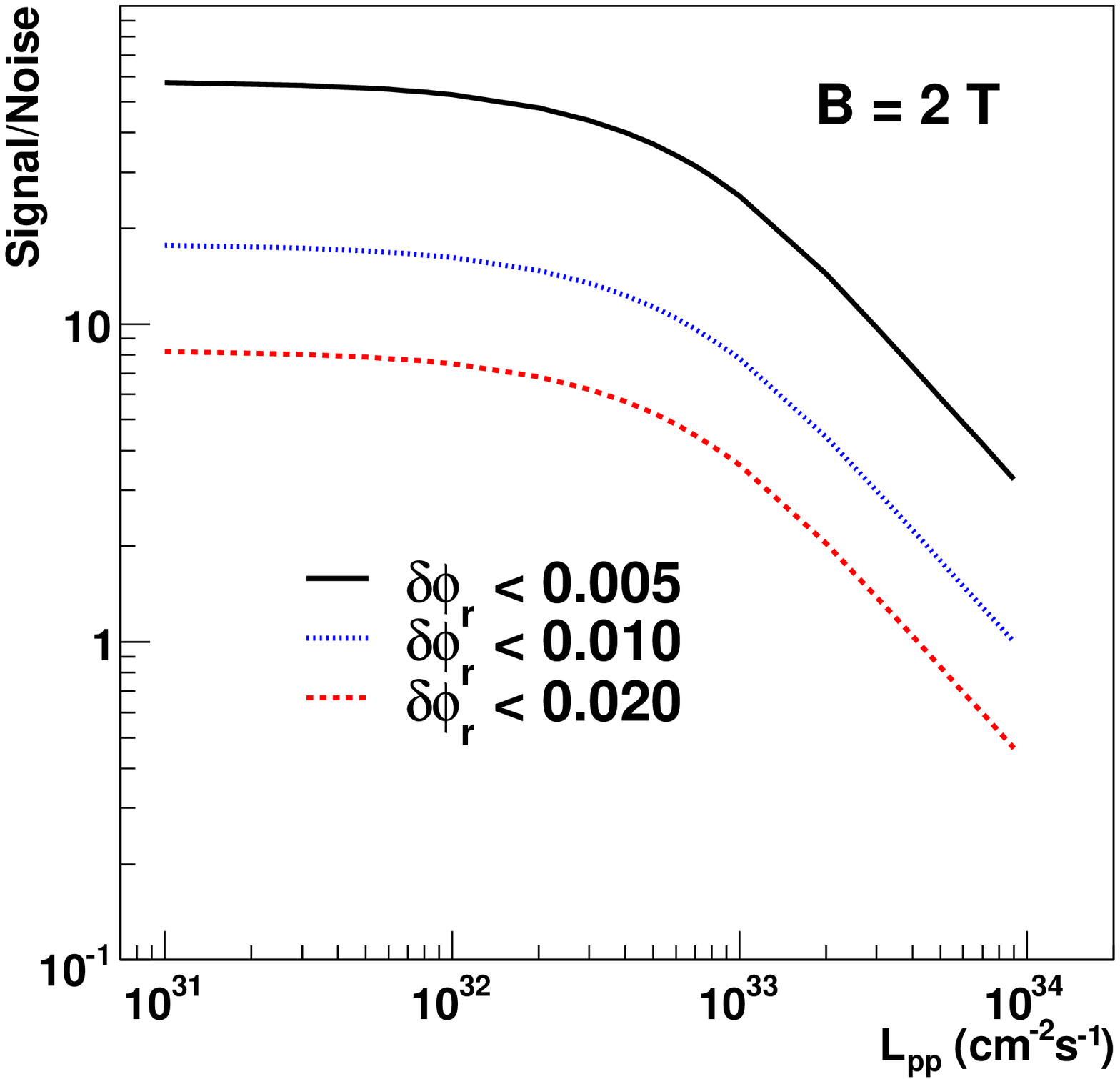, width=65mm,height=65mm}
}}
\put( 33,-5){\makebox(0,0)[cb]{\bf (a)}}
\put(95,-5){\makebox(0,0)[cb]{\bf (b)}}
\end{picture}
\end{center}
\caption{The signal to noise ratio for the three values of the acoplanarity cut
as a function of the machine luminosity:
--~(a)~for the $B = 0$~Tesla field configuration,
--~(b)~for the solenoidal field of $B = 2$~Tesla. }
\label{plot18}
\end{figure}

In Figure \ref{plot18} the ratio of the signal to the background rates at the EF 
selection stage is shown as a function of the machine luminosity for the three 
values of the upper limit of the reduced acoplanarity for the two settings of 
the detector magnetic field. These plots show that, already for the model of the 
base-line detector, the proposed event selection procedure assures a comfortable 
value of the signal to noise ratio over a large range of the machine 
luminosities. The drop of the signal to noise ratio for large luminosities is 
driven solely by a decreasing probability of the silent bunch crossing. For the 
$\bf{B2}$ case the signal to background ratio can be improved significantly by 
selecting events in a narrow bin of the reduced acoplanarity.

 \begin{figure}
\begin{center}
\setlength{\unitlength}{1mm}
\begin{picture}(130,60)
\put(0,0){\makebox(0,0)[lb]{
\epsfig{file=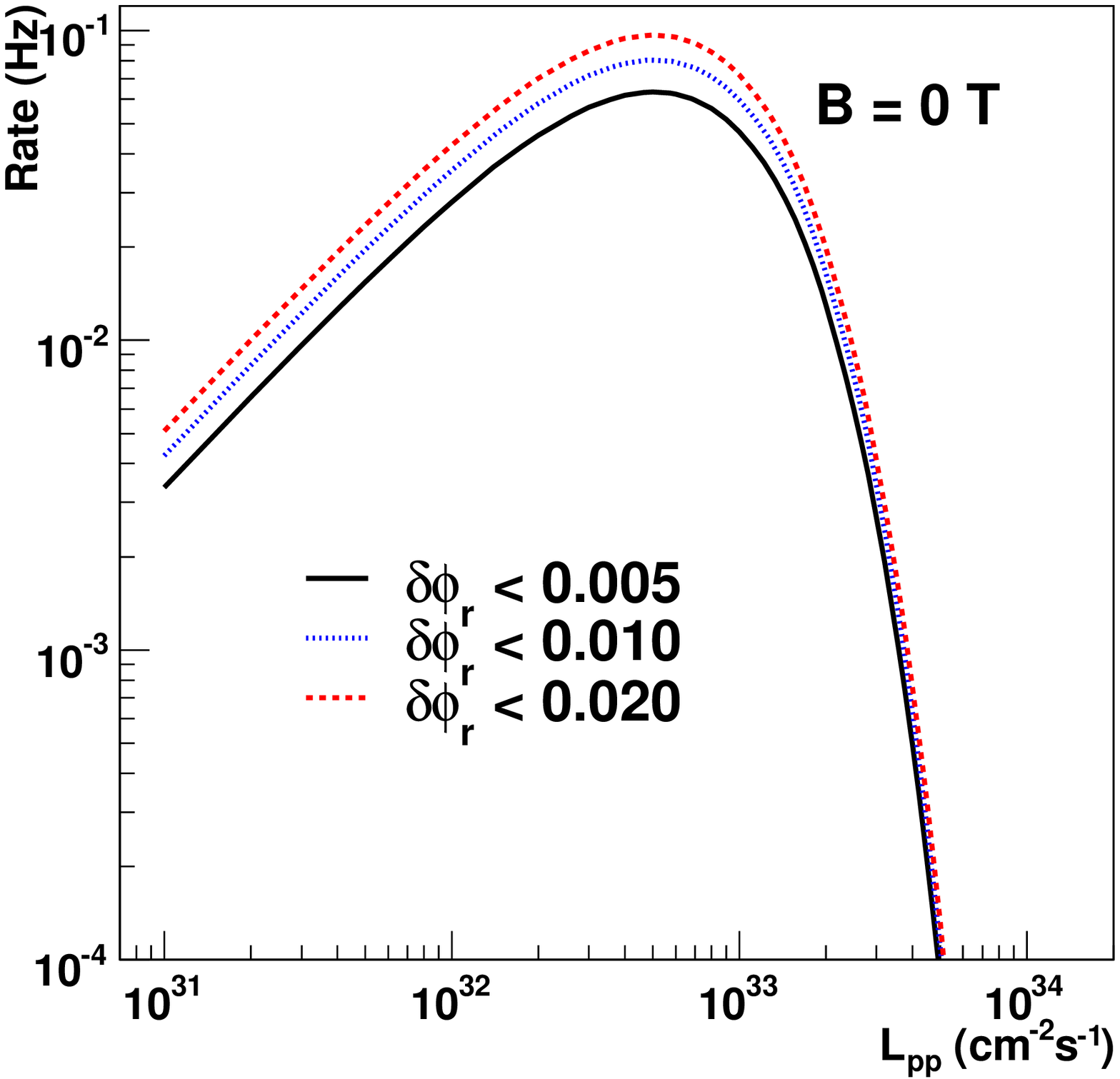, width=65mm,height=65mm}
}}
\put(65,0){\makebox(0,0)[lb]{
\epsfig{file=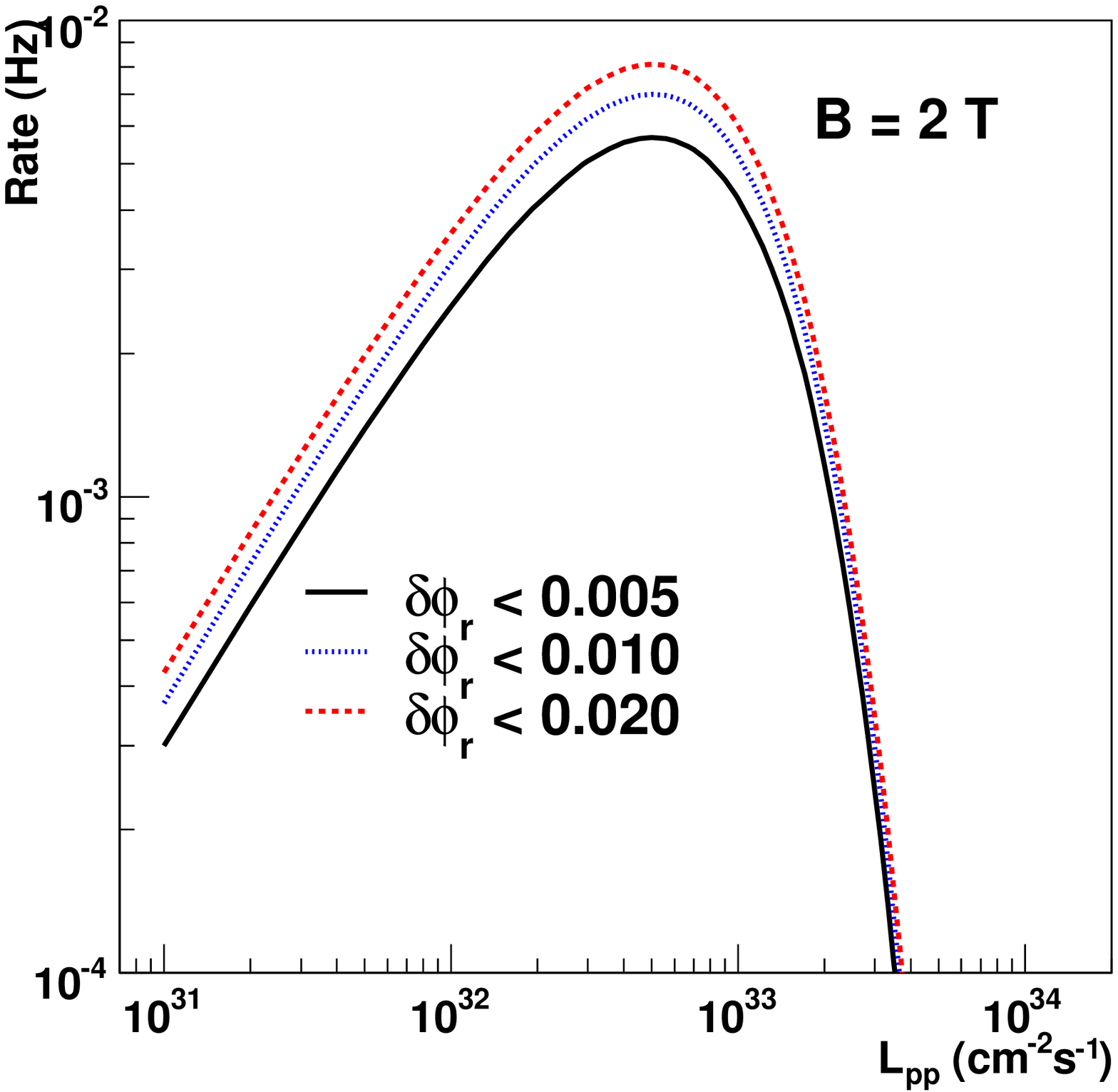, width=65mm,height=65mm}
}}
\put( 33,-5){\makebox(0,0)[cb]{\bf (a)}}
\put(95,-5){\makebox(0,0)[cb]{\bf (b)}}
\end{picture}
\end{center}
\caption{The signal event rate for the three values of the acoplanarity cut
as a function of the machine luminosity: --~(a)~for the $B = 0$ Tesla field 
configuration, --~(b)~for the solenoidal field of $B = 2$ Tesla.}
\label{plot19}
\end{figure}

In Figure \ref{plot19} the signal rates for events passing the EF selection 
stage are shown as a function of the machine luminosity for the three values of 
the reduced acoplanarity cut. The rates drop significantly below the initial 
level of 1~Hz level discussed in \cite{first} but are sufficiently high to 
provide a statistically precise measurement of the luminosity. 

For the luminosities below $L\simeq 6\cdot 10^{32}$~s$^{-1}$cm$^{-2}$ the rate 
of the signal events rises with increasing luminosity. At higher luminosities 
the apparent rate drop is a consequence  increasing average number of collisions 
per bunch crossing. For $L \simeq 6 \cdot 10^{32}$s$^{-1}$cm$^{-2}$ a 1\% 
statistical precision can be achieved over the integrated time intervals of 
about 30~hours for the $\bf{B0}$ configuration and about 400 hours for the 
$\bf{B2}$ configuration. These time intervals could be decreased significantly 
if the LVL1 track acoplanarity tagging were made in vicinity of the collision 
vertex - a solution which is presently out of reach for the existing trackers of 
the LHC detectors but can be reconsidered while upgrading the LHC detector's 
trackers for the high luminosity phase of collider operation.

It is important to note that the direct extension of the proposed method beyond 
the luminosity value of $L= 10^{33}$s$^{-1}$cm$^{-2}$ requires at least one of 
the following three possible upgrades  of the base-line luminosity detector:  
\begin{itemize}
\item adding a precise hit-timing measurement,  
\item adding the z-planes to the luminosity detector and developing fast 
algorithms capable to determine the z-position of the track origins with a 1~mm 
precision within the LVL1 trigger latency,
\item  adding the LVL1 trigger electron/pion rejection capacity functions.   
\end{itemize}

If such upgrades are not made, the absolute luminosity can be measured in the 
low and medium luminosity periods and subsequently ``transported" to the high 
luminosity periods. Foundation of such a procedure is precise monitoring of the 
relative instantaneous luminosity over the whole range of the LHC luminosities 
and in fine time intervals. 

\section{Monitoring of Instantaneous Luminosity }  

\subsection{Goals}

There are two basic reasons to measure the relative, instantaneous changes of 
the luminosity, $L(t)$, with the highest precision:
\begin{itemize}
\item the method discussed here can reach the absolute luminosity  precision 
target only for  selected runs and selected bunch crossings and must be 
subsequently transported (extrapolated) to  all the runs and bunch crossings,
\item the statistically significant samples of the electron-positron pair 
production events are  collected over the time periods which are sizeably 
longer than the time scales of the changes in the detector recorded data quality 
(the electronic noise, the beam related noise, the event pile-up, the detector 
calibration and efficiencies, etc.) -- the time evolution of the corresponding 
corrections must thus be weighed according to the instantaneous luminosity in 
the precision measurement procedures. 
\end{itemize} 

Precise measurement of the relative instantaneous luminosity in fine time 
intervals is bound to be based on the rate of strong rather than electromagnetic 
interactions of the beam particles. In the scheme proposed in this paper the 
on-line luminosity is determined by counting the luminosity detector in-time 
track segments produced in strong interactions of the colliding protons. It is 
determined using solely the luminosity detector data and its local data 
acquisition system. The on-line luminosity is determined in ${\cal O}(1)$~minute  
intervals. The corresponding final off-line luminosity is then recalculated  
using the data recorded by the host detector. 

\subsection{Counters}

The following luminosity detector counters are proposed for a 
precision measurement of the instantaneous, relative on-line luminosities:  

\begin{enumerate} 

\item {\bf Track-Global} - representing the mean number of the in-time track 
segments per bunch crossing both in the left and in the right side of the 
luminosity detector,

\item {\bf Track-Global-OR} - representing the mean number of the in-time track 
segments per bunch crossing seen in both parts of the luminosity detector with 
the additional requirement that only those bunch crossings are considered  for 
which there is at least one in-time track segment in either the left or in the 
right side of the luminosity detector,

\item {\bf Track-Global-AND} - representing the mean number of the in-time 
track segments per bunch crossing in both sides of the luminosity detector with 
the additional requirement that only those bunch crossing are considered for 
which there is at least one in-time track segment in each of the luminosity 
detector sides,

\item {\bf Track-Event-OR} - representing the fraction of bunch crossings with 
at least one in-time track segment either in the left or in the right half of 
the detector,

\item {\bf Track-Event-AND} - representing the fraction of bunch crossings with 
at least one in-time track segment 
in each of the detector halfs,

\item {\bf Track-Left (Right)}  - representing the mean number of the in-time 
track segments per bunch crossing in the left (right) part of the luminosity 
detector,

\item {\bf Track-Event-OR-Left(Right)} -  representing the fraction of bunch 
crossings with at least one in-time track segment in the left (right) side of 
the luminosity detector,

\item {\bf Track-Sector($i_{L(R)})$} - representing the mean number of the 
in-time track segments per bunch crossing in the $i_{L(R)}$--th $\phi$-sector of 
the left (right) side of the luminosity detector,

\item {\bf Track-Sector-OR($i_{L(R)})$} - representing the fraction of bunch 
crossings with at least one in-time track segment in the $i_{L(R)}$--th
$\phi$-sector of the left (right) side of the luminosity detector,

\item {\bf Track-Sector-Coinc($i_L,i_R)$} - representing the mean number  of the 
in-time track segments per bunch crossing in the $i_{L}$--th $\phi$-sector of 
the left detector side and in the $i_{R}$--th $\phi$-sector of the right 
detector side,
 
\item {\bf Track-Sector-Coinc-AND($i_L,i_R)$} - representing the fraction of 
bunch crossings with at least one in-time track segment both in the $i_{L}$--th
$\phi$-sector of the left detector side and in the $i_{R}$--th $\phi$-sector of 
the right detector side,

\item {\bf Track-SBC} - representing the fraction of bunch crossings with no 
in-time track segments in each detector side. 

\end{enumerate}

Subdividing the sample of track segments into 36 $\phi$-segment sub-samples 
allows to provide precise measurements over the whole range of the LHC 
luminosities ($10^{30} - 10^{34}$~cm$^{-2}$s$^{-1}$) independently of the 
number of pile-up collisions occurring within the same bunch crossing.

The counting is done separately  for paired, unpaired (isolated and 
non-isolated) and empty (isolated and non-isolated) bunch crossings. The 
counters proposed above, together with the corresponding hit-based and 
out-of-time track segments based counters, provide the necessary input data 
to determine also the instantaneous luminosity of the LHC in the whole range of 
the average number of collisions per bunch crossing, $< \mu >$ and in the  full 
range of the dispersion of the bunch-by-bunch luminosity. The track segment 
based counters, contrary to the LUCID or BCM detector ones \cite{ATLAS_LUMI}, 
are insensitive to the beam induced background and afterglow effects obscuring 
the extrapolation of the van der Meer scan luminosity to arbitrary data 
collection periods. Moreover, they can be precisely controlled in the off-line 
analysis of the host detector tracks traversing the luminosity detector volume.

While the first five counters provide a fast diagnostic for the $L$ and $<\mu >$ 
dependent optimisation of the luminosity counting method, the following six 
counters are used directly by the instantaneous luminosity measurement 
algorithms. A detailed presentation of the counting algorithms is outside the 
scope of the present paper and will not be discussed here. The only aspect 
which may be elucidated is the extension of presently applied methods of the 
instantaneous luminosity measurement \cite{ATLAS_LUMI} to the full range of $< 
\mu >$. As the $< \mu >$ value increases the inclusive track counters are 
replaced, at first by counting of tracks separately in each of the 
$\phi$-sectors of both detector parts, and eventually, at the highest $< \mu >$, 
by counting of the left-right $\phi$-sector coincidences. For such a  
``step-by-step" procedure the unfolding of the number of interaction per bunch 
crossing $< \mu >$ is no longer necessary. 

Since the $L(t)$ value is determined locally by the luminosity detector 
algorithms it is, by definition, independent of the host detector dead time. 
Moreover, the relative luminosity can be monitored over the time periods when 
the host detector sub-components are in the stand-by mode. 

The precise off-line corrected $L(t)$ values can be determined for all the
time periods for which at least the tracker and the calorimeter are in the 
data taking mode. The off-line correction factors can  be determined using the 
sample of the host detector reconstructed tracks traversing the fiducial volume 
of the luminosity detector. The tracks are parasitically sampled with 
$\cal{O}$(1)~kHz frequency using the sample of the host detector recorded 
events. 

In the method based on the track segments the event pile-up plays a positive  
role. It allows to increase the track sampling frequency and, as a consequence,  
to control the off-line correction factors in finer than ${\cal O}(1)$~minute  
time intervals.
 
\section{Absolute Luminosity}

\subsection{Low Luminosity Periods}

We shall consider first the case of the luminosity determination during the low 
instantaneous luminosity periods defined by the following condition on the 
average number of interaction per bunch crossing: $<\mu> \ll 1$. In these 
periods the probability of the silent bunch crossings is sufficiently large to 
base the luminosity determination on the measurement of the rate of the bunch 
crossings with exclusive production of electron-positron  pairs. 

The integrated luminosity $L_{int}$ is calculated using the following formula:
\begin{equation}
L_{int} = \sum_{t_i} \frac{N_s(t_i)\cdot(1-  \beta(t_i))}{P^{silent} (t_i) \cdot  Acc(t_i) \cdot \epsilon(t_i) \cdot \sigma_{e+e-} }
\label{eq:lumi}
\end{equation} 
where, 
\begin{itemize}
\item $N_s(t_i)$ is the total number of the exclusive electron-positron pair 
candidates passing the LVL1, LVL2 and EF selection criteria which were recorded 
over the time interval $(t_i, t_i +\Delta t_i)$;

\item $\beta(t_i)$ is the fraction of the total number of the exclusive 
electron-positron pair candidates passing the LVL1, LVL2 and EF selection 
criteria which originate from the background strong interaction processes. This 
quantity is determined using a monitoring sample of the reconstructed TPZ 
trigger events. The rate of pairs created in strong interaction is measured in 
the $0.1 < \delta\phi_r < 0.3 $, where the contribution of the electromagnetic 
processes is negligible, and subsequently extrapolated to the signal  
$\delta\phi_r  < \delta\phi_r^{cut}$ region. This extrapolation is insensitive 
to the particle production mechanism in strong interactions and can be performed 
in the model independent way. It is important to note that, as far as the silent 
bunch crossings are concerned, the time variation of $\beta(t_i)$ is very week 
as compared to the time variation of $N_s(t_i)$ and $P^{silent}(t_i)$. As a 
consequence only insignificant increase of the fraction of the host detector 
LVL1 band-width is required for the TPZ accepted events. The background sources 
other than those related to the genuine strong interactions processes are 
controlled using the unpaired and empty bunch crossings;

\item $P^{silent} (t_i)$ is defined as:
$$P^{silent} (t_i) = \frac{R_{SBC}}{R_{BC}},$$ 
where $R_{BC}$ and $R_{SBC}$ are, respectively, the total number of paired 
bunch crossings and the total number of silent bunch crossings in the sample of 
paired bunch crossings within the time interval $(t_i, t_i +\Delta t_i)$; 

\item $Acc(t_i)$ is the acceptance for the electron-positron pairs traversing 
the luminosity detector and  satisfying the $ \delta\phi_r < \delta\phi_r^{cut}$ 
condition. The acceptance correction includes the detector smearing effects,
the geometric acceptance of the luminosity detector, and all the dead material 
effects. The $Acc(t_i)$ values are determined in the model independent way using 
those of the particles produced are in recorded strong interaction collisions 
which traverse both the luminosity detector and the host detector tracker. The 
momentum scale, the detector smearing and the dead material effects, discussed 
in \cite{second}, are directly measured using the abundant sources of electrons 
and positrons -- the conversions of photons coming from the decays of neutral 
pions in the material of the beam pipe. The correction factors sensitive 
to the precise position of the electron (positron) track origin, are determined 
using the electron-positron pairs from Dalitz decays. The time variation of the 
acceptance due to an increase of the longitudinal emittance of the proton beam 
over the LHC run is controlled using the time evolution of the $z$-vertex  
distribution for the bulk of recorded events;  

\item the efficiency $\epsilon(t_i)$ can be decomposed as follows: 
$$\epsilon(t_i) = \epsilon_{extr-}(t_i) \cdot\epsilon_{extr+}(t_i) 
\cdot\epsilon_{id-}(t_i) \cdot\epsilon_{id+}(t_i)\cdot P^{silent}_{LVL2/EF} 
(t_i),$$
where: $\epsilon_{extr+}(t_i) (\epsilon_{extr-}(t_i))$ is the efficiency of 
linking of the positive (negative) charge luminosity detector track segments to 
the vertex constrained, SCT/Pixel ones; $\epsilon_{id-}(t_i)$ 
$(\epsilon_{id+}(t_i))$ is the electron (positron) identification efficiency in 
the host detector LAr calorimeter; $P^{silent}_{LVL2/EF} (t_i)$ is the fraction 
of the luminosity detector silent bunch crossing with no  LUCID (BCM) particle 
hits and no reconstructed charged particle tracks pointing the electron-positron 
pair vertex. The linking efficiency and the electron/positron identification 
efficiencies are determined using the full sample of recorded and reconstructed 
events. Their rate (${\cal O}(200)$~Hz) is sufficiently large for a precise 
control of the time dependence of these efficiencies. The 
$P^{silent}_{LVL2/EF}(t_i)$ values are determined using the CTP-prescaled 
fraction of the SBC triggered events. As in the previous case, the time 
variation of $P^{silent}_{LVL2/EF}(t_i)$ is by far less important than that of 
$N_s(t_i)$ or $P^{silent}(t_i)$; 

\item $\sigma_{e+e-}$ is the total exclusive $e^+e^-$ pair production cross 
section. For exclusive coplanar pairs reconstructed in the fiducial volume of 
the luminosity detector this cross section is largely dominated by  the cross 
section for peripheral collisions of the beam particles mediated by two photons 
\cite{first}. 
\end{itemize} 

\noindent
The acceptance and efficiencies depend upon the momenta of the electron and 
positron. This dependence and the corresponding integrations in eq. 
(\ref{eq:lumi}) was dropped in the formulae for simplicity.  

The strength of presented method is that it is based on low $p_T$ 
electrons/positrons which are produced abundantly in the minimum bias collisions 
and recorded with the host detector, independently of the LVL1/LVL2 and EF class 
of events. These particles play the role of high precision calibration candles 
for the luminosity measurement procedures allowing to avoid almost completely 
the use of the Monte-Carlo based methods relying both on the modeling of the 
strong interactions and on the modeling of the luminosity detector performance. 
In addition, since the luminosity events are processed by the TDAQ system of the 
host detector, no corrections for the detector dead time and event losses at 
various stages of the data filtering process are required for the absolute 
normalisation of any recorded data sample.  

\subsection{Medium Luminosity Periods} 

In the phases of the LHC operation when the average number of $pp$ interactions 
per bunch crossing is $<\mu> = {\cal O}(1)$ the probability of an overlap of the
electromagnetic and the strong interaction driven collisions becomes large and 
the losses of the $e^+e^-$ pair events by applying the exclusivity criteria at 
the LVL2 and EF levels need to be monitored with significantly higher precision 
and at much finer time intervals. 

The remedy is to extend the definition of the Silent Bunch Crossing, based so 
far exclusively on the luminosity detector signals, to a Global Silent Bunch 
Crossing (GSBC) based on the CTP  coincidence of the SBC bit with the 
corresponding SBC bits coming from the LUCID and from the BCM detectors. 
The GSBC occurrence probability, $P_G^{silent}(t_i)$, would have to be monitored 
with precision similar to that of the instantaneous luminosity. 

Another, more elegant solution, is to multiplex selected LUCID and BCM LVL1 
trigger signals and to send them as the input signals to the luminosity detector 
trigger logic. In this case, a care would have to be taken to position the 
luminosity detector trigger electronics racks in a place where the LUCID and BCM 
signals could arrive in-time. 

The luminosity formula  \ref{eq:lumi} remain valid for the medium luminosity periods. The only 
change with respect to the low luminosity case is to replace $P^{silent}(t_i)$ 
by $P_G^{silent}(t_i)$ and to replace $P^{silent}_{LVL2/EF}(t_i)$ by 
$P^{silent}_{EF}(t_i)$ representing the fraction of the global silent bunch 
crossings in which no reconstructed charged particle tracks pointing the lepton 
pair vertex were found within the tracker volume.

\subsection{High Luminosity Periods With Base-line Luminosity Detector} 

For the phases of the LHC operation when the average number of $pp$ interactions 
per bunch crossing $<\mu> \gg 1$ the method based on the counting of $e^+e^-$
pairs in the restricted sub-sample of silent bunch crossings does not work any 
longer. There are two ways to proceed. An optimal one is to upgrade the 
capacities of the luminosity detector. This will be discussed in the next 
section. Another one, discussed below, uses the base-line detector and 
reorganises the data taking at the expense of a small reduction of the 
time-integrated luminosity. This method uses only a fraction of collisions for 
which the condition $<\mu> = {\cal O}(1)$ is fulfilled. The absolute 
luminosity is then extrapolated to the arbitrary   data taking period using the 
measurement of the relative, instantaneous luminosity.

There are three ways of collecting  $<\mu> = {\cal O}$(1) data over the high 
luminosity running period of the LHC. 
\begin{enumerate}

\item The first, obvious one, is based on dedicated machine runs with reduced 
luminosity per bunch crossing. If a fraction below 10\% of the machine running 
time is devoted to such runs the effect of reduced overall luminosity on the 
physics results would  be unnoticeable for the searches of rare events and very 
useful for the physics programs requiring large samples of events with single 
collisions per bunch crossing. This programme profits from the relatively large 
cross section for the electron-positron pair production.
 
\item In the second method the luminosity detector triggers are activated only 
at the end of the machine luminosity run when the currents of the beams decrease 
or the beam emittance increases such that the $<\mu> = {\cal O}(1)$ condition 
is fulfilled. The applicability of this method depends upon the beam life-time 
and the run-length of the collider. For the present running strategy, maximising 
the collected luminosity, only a small increase of the range is feasible. 

\item The third method would require a special LHC bunch-train injection pattern 
in which one of the twelve bunch trains (4 $\times$ 72 bunches), reflecting the 
complete Linac, Proton Synchrotron Booster (PSB),  Proton Synchrotron (PS) and 
Super Proton Synchrotron (SPS) cycle, contains bunches with a reduced number of 
protons. The reduction factor, $~ \sqrt{<\mu>}$, depends upon the average 
number of collisions per bunch crossing for the remaining eleven bunch trains.  
The luminosity detector triggers are proposed to be masked unless they are in 
coincidence with crossings of the low intensity bunches. If such running mode 
can be realised at the LHC\footnote{Such a running mode is anything but easy in 
the presence of coherent bunch interaction effects which depend on the bunch 
charge. The maximal acceptable dispersion range of the bunch intensities would 
have to be determined by the LHC machine experts. There  are several 
consequences for such a running scenario. For example, one has to take into 
account here that the four LHC experiments are running at the same time, and 
that the colliding bunch partners are different in different Interaction Points 
(IPs) of the LHC -- it is worth stressing here that the distance between the 
ATLAS and CMS IPs is half of the LHC ring circumference, and that the IPs at 
which the bunches of different intensity interact are those of the ALICE and 
LHCb experiments for which maximising the collected luminosity is of secondary 
importance.} the absolute and relative luminosities could be sampled over the 
same time periods. This could allow for a drastic reduction of all the relative, 
time dependent measurement uncertainties. More importantly, a concurrent 
storage of the highest possible intensity bunches with the low intensity bunches 
at the LHC would be beneficial for the LHC precision measurement programme. It 
would allow concurrent measurement of the pile-up effects in those of the 
physics observables which need to be measured over the large time span, thus 
inevitably over a large $<\mu>$ range. The above running scenario is technically 
feasible \cite{Myers}, but requires a wide consensus of the four LHC 
experiments. 

\end{enumerate}

In each of the above strategies the extrapolation of the absolute luminosity  
measured for the $<\mu>= {\cal O}(1)$ bunch crossings, to an arbitrary bunch 
crossing set and data collection period is derived from the following 
$\phi$-sector track counters: {\bf Track-Sector$(i_{L(R)})$}, 
{\bf Track-Sector-OR$(i_{L(R)})$}, {\bf Track-Sector-Coinc$(i_L,j_R)$} and 
{\bf Track-Sector-Coinc-AND$(i_L,j_R)$}). The reason for choosing the method 
based upon the $\phi$-sector track rates is to assure that the probability of an 
observation of a track segment in a restricted phase space per minimum bias 
event is sufficiently small to disregard the pile-up effects in the luminosity 
calculation algorithms in the whole luminosity range: 
({\bf Track-Sector($i_{L(R)}$)}$ \ll 1$ or {\bf Track-Sector-Coinc($i_L,j_R)$} 
$\ll 1$). The statistical precision of this method is assured by the use of the 
mean values over all the $\phi$-sectors of the {\bf Track-Sector-OR($i_{L(R)})$} 
or {\bf Track-Sector-Coinc-AND($i_L,j_R)$}. 

\subsection{High Luminosity Periods with Upgraded Luminosity Detector} 

In all the studies presented so far in this paper the model of the base-line
detector was used. For the direct measurement of the  absolute luminosity in 
large $<\mu>$ runs an upgrade of the detector functionalities is necessary. The 
detailed discussion of such an upgrade is beyond the present work scope. 
However, it is worth sketching already here the basic conceptual and hardware 
aspects of such an upgrade. 

The principal upgrade goal is to provide, within the LVL1 trigger latency, not 
only the luminosity detector in-time track segments but, in addition, the 
measured $z$-positions of their origin with ${\cal O}(1)$mm precision. The 
search of the coplanar particle pairs must be restricted, in large $<\mu>$ runs 
producing multiple vertices, only to the track segments pointing to the same 
vertex and satisfying the condition that no other luminosity detector track 
segment, except for the two coplanar tracks, is associated with it. Similarly, 
at the EF level, the luminosity events have to be selected only if there were no 
reconstructed host detector tracks pointing to the $e^+e^-$ pair vertex. It 
should be noted that the LUCID and the BCM detectors' signals can no longer be 
used in the search process of the exclusive electron-positron pairs. The 
corresponding reduction of the rejection power of the background events based on 
the exclusivity criteria would have to be recuperated by a more efficient  
electron/pion recognition. On the other hand, the restriction of the luminosity 
measurement to only the silent bunch crossings would no longer be necessary. The 
rate of selected electron-positron pairs would increase significantly at the 
price of a smaller signal to noise ratio, which in turn could be compensated by 
a higher cut-off on the electron (positron) momentum leading to a better pion 
rejection than the one assumed in the base-line model.  

Two directions  of upgrading the luminosity detector can be singled out. The 
first one is to use the detector technology which provides a high precision 
timing of the particle hits, such as the one being developed for the Roman Pot 
Project \cite{RomanPot}. The use of the hit timing was discussed in our earlier 
paper \cite{second}. In addition an extra  radial segmentation of the luminosity 
detector would have to complement the hit-timing based backtracking  by 
reconstructing the track segments in three rather than in two dimensions. Both 
can be achieve applying for example the micromegas technology \cite{micromegas}. 
 
The second one, is to use a hadron blind detector \cite{Charpak}. Note that in 
both cases a significant increase of the processing power of the LVL1 trigger 
FPGAs would be required. 

\section{Merits of  {\bf B0} Runs} 

All the algorithms and methods presented above can be used for the runs with 
the nominal strength  of the host detector magnetic field and in those for which  
the solenoid magnetic field switched off. 

The merits of supplementing the standard {\bf B2} configuration runs by the
dedicated, {\bf B0} ones are numerous. First of all, the luminosity measurement 
becomes almost insensitive to the radiation of photons by the electrons and 
positrons traversing the host detector dead material. The {\bf B0} runs allow to 
cross-check the precision of the understanding of radiation effects which 
otherwise have to be monitored  using the Dalitz pairs and photon conversions. 
Moreover, the complexity of the luminosity detector FPGA algorithms would be 
drastically reduced facilitating the applicability  of the proposed method to 
the periods of high luminosity. This is because, only in the {\bf B0} case, the 
algorithms reconstructing the vertices of the interactions occurring in the 
same bunch crossing can be decoupled from the acoplanarity algorithms. In 
addition, the EF exclusivity cut based on the reconstructed host detector 
reconstructed tracks is significantly more efficient in rejecting fake 
electron-positron pairs due to the strong interactions (all the charged particle 
tracks including  a very low momentum ones could be reconstructed). It has to be 
stressed that the rate of the selected coplanar electron-positron  pairs in the 
high luminosity periods reaches the ${\cal O}(1)$Hz level. Therefore, only 
insignificant fraction of the luminosity would be sacrificed for the absolute 
luminosity measurement runs. 

The propagation of the absolute luminosity from the {\bf B0} runs to the {\bf 
B2} runs can be performed precisely and elegantly by a simultaneous measurement 
of the rate $Z$ bosons in the {\bf B0} runs\footnote{The invariant mass of the 
electron-positron pair is determined in the {\bf B0} runs using the angles of 
the leptons and the LAr energy deposits linked to the lepton tracks.} which can 
be subsequently used for the absolute luminosity extrapolation to any {\bf B2} 
period of the duration larger than that required to collect ${\cal 
O}(10)$pb$^{-1}$ of the integrated luminosity. Note, that several systematic 
uncertainties of the absolute luminosity measurement cancel in the ratio of 
cross section for the coplanar electron-positron pair production and to that for
the $Z$ boson production (among these are the LAr absolute scale error, 
remaining effects of electron radiation, etc..,) 

The advantage of the scheme described above comes from the extraordinary
coincidence that in the fiducial volume of the proposed luminosity detector the 
rate of the electron-positron pairs produced in ``elastic" collisions of 
point-like protons is comparable to the rate of the $Z$ bosons produced in the 
inelastic ones. 
 
\section{Absolute Cross Sections}

The luminosity method proposed in this series of papers extends and simplifies 
the present techniques of the absolute normalisation of those of the 
distributions of physical observables which are derived from the  data collected 
over a long time-period. The standard luminosity block based technique 
\cite{ATLAS_LUMI} remains valid. However, it is no longer indispensable.  

In the proposed method users could choose independently the optimal fraction of 
collected data satisfying  the data quality criteria defined on the  
bunch-by-bunch rather than on the block-by-block basis. The only requirement is 
that the same criteria are used both for the user selected sample and for the 
corresponding sample of the electron-positron pair luminosity  events. Instead 
of the list of valid luminosity blocks the user would  be provided with the 
ready-to-use offline algorithms to calculate the integrated luminosity 
corresponding to the user-specific data quality criteria\footnote{Physicists 
using the data coming from different detector components may prefer different 
data quality criteria. In addition, depending on their tasks they may reject 
algorithmically, again on bunch-by-bunch basis, a sample of selected bunch 
crossings properties e.g. reject the bunch crossings with interactions of the 
halo particles within the detector volume.}. The overall detector dead time and 
all the losses of the events at any stage of the data selection process which 
are independent of the data content (LVL2/EF processors' timeouts, etc.) need no 
longer be monitored and accounted for - these effects are automatically taken 
into account in the coherent analysis of the two data samples: 
\begin{itemize}
\item the user selected sample, 
\item the electron-positron pair luminosity event sample.  
\end{itemize}

The key simplification, proposed below, is to avoid altogether all the 
uncertainties on the rate of the silent bunch crossing and on the time 
dependent efficiencies of the $e^+e^-$ pair selection. The underlying trick is 
to assign each event of the user selected samples of bunch crossings into the 
following two samples: the first one containing all the user preferred events 
and the second one containing only those events for which there were no 
luminosity detector track segments other than those associated with selected 
$pp$ collision. The absolute luminosity is determined first for the latter class 
of events and subsequently extrapolated, using the measurement of the relative 
luminosity, to the full sample. For such a procedure the uncertainties in the 
monitoring of the silent bunch crossings cancel in the ratio of the 
numbers of accepted luminosity and the user-selected events. The 
electron-positron pair selection efficiencies  for the luminosity events are  
calculated, within this method, using only the soft particles produced in the 
user-selected  events. Thus all the efficiencies can  be sampled in the same way
for the luminosity and for the user selected events. Precise monitoring of 
the time dependent detector and the beam quality related features is thus no 
longer necessary.  

\section{Systematic Uncertainties}  

The overall luminosity measurement errors, are  dominated by the systematic 
measurement uncertainties\footnote{ For example, for the machine luminosity of 
$L= 10^{33}$s$^{-1}$cm$^{-2}$ the sampling time to reach the 1 \% statistical 
precision of the luminosity measurement is three days for the {\bf B0} runs and 
one month for the {\bf B2} runs.}. The key element of the measurement strategy 
presented in this series of papers, allowing for a significant reduction of the 
systematic errors and for a precise control of the remaining ones, is the 
placement of the luminosity detector within the host detector tracker and 
calorimeter geometrical acceptance regions.

The dominant systematic uncertainties, reflecting all the time dependent aspects 
of the machine and detector performances, can be controlled using the soft 
particle tracks and energy deposits in the sample of ${\cal O}(100)$ bunch 
crossings recorded by the host detector. The rate of soft particles is large 
enough to use only the preselected bunch crossing data for the 
monitoring\footnote{Of the order of 10$^{5}$ electrons (positrons) coming from 
the photon conversions and crossing the luminosity detector volume are recorded 
over one minute.}. If the sample of the bunch crossings chosen for the 
monitoring is identical to the one selected for the measurement of any given 
observable than, by definition, both the luminosity measurement monitoring data 
and the user selected data are sampled concurrently. Such a concurrent sampling 
does not only reduce the systematic errors but, in addition, simplifies the 
procedure of the absolute normalisation of the measured observables by replacing 
the time dependent quantities, entering the luminosity master formula, by the 
quantities averaged over the selected sample of bunch crossings. This is one of 
the merits of the proposed method. It allows for a continuous improvement of the 
accuracy of the absolute measurements with the increase of the measurement time 
without a necessity of a precise book-keeping of all the time dependent detector 
and machine performance features.  

\subsection{Background Subtraction}

\begin{figure}[h]
\begin{center}
\epsfig{file=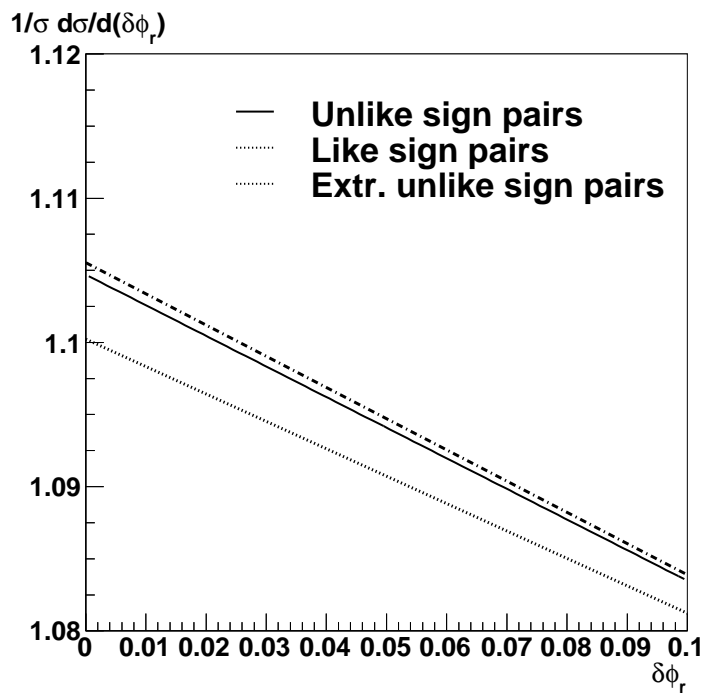,width=10cm}
\end{center}
\caption{The estimation of the upper bound of the background subtraction 
systematic uncertainty. See text for explanation}
\label{plot20}
\end{figure}

The fraction of the total number of exclusive electron-positron pair candidates 
coming from the strong interaction background sources, $\beta(t_i)$, is  
determined using solely the monitoring data, bypassing all the modeling 
uncertainties of the minimum bias events. The upper bound of its initial 
(anticipated) systematic uncertainty was determined using the simulated PYTHIA 
events. This procedure, illustrated in Figure~\ref{plot20}, mimics 
algorithmically the ultimate procedure based on the recorded events.
 
The continuous line, depicted in Figure~\ref{plot20}, represents a linear fit to 
the reduced acoplanarity distribution of the unlike charge particle pairs for 
those of the background events (PYTHIA) for which the TPZ LVL1 trigger bit was 
set to 1. This distribution was fitted in the $(0.1, 0.3)$ interval in which the 
contribution of the electron-positron pair events is negligible \cite{first}, 
and then extrapolated down to $\delta\phi_r = 0$ using the fit parameters. The 
extrapolation result is represented  by the dash-doted line in 
Figure~\ref{plot20}.
 
In the next step, the $\delta\phi_r$ distribution for the like charge pairs
satisfying the $TPZ = 1$ condition was fitted in the $(0.1, 0.3)$ interval and 
extrapolated to $\delta\phi_r = 0$ using the fit parameters. The extrapolation 
result is marked with the dotted line in Figure~\ref{plot20}. The upper bound of 
the background subtraction uncertainty corresponds to the spread of the 
distributions and is of the order of 0.4\% for $\delta\phi_r < 0.05$\footnote{ 
Note that the pairs used in the luminosity measurement contribute to the like 
sign pair sample only if the charge of one of the particles charge is wrongly 
reconstructed. For the low momentum particles the probability of the particle 
charge misidentification is small enough to be neglected. The sample of like 
sign pairs represents, thus, a pure background sample.}.  
This plot proves that the extrapolation to the small acoplanarity region is 
insensitive to the total particle pair charge and reflects merely the 
phase-space for multi-particle production in strong interactions.

\subsection{Acceptance Correction}

The systematic uncertainties  on the acceptance, $Acc(t_i)$, are, as before, 
controlled using the host detector recorded  data. The systematic effects are 
subdivided into two classes: the host detector effects and the luminosity 
detector effects.  

The contribution of the host detector effects such as:
\begin{itemize}

\item the losses of the electrons and positrons on the way from the interaction 
vertex to the luminosity detector fiducial volume -- due to hard photon 
radiation in the material of the beam pipe or of the host detector,  

\item the biases in the reconstructed momentum scale of those of the host 
detector reconstructed charged particles which traverse the fiducial volume of 
luminosity detector, 

\item the momentum resolution biases, 

\item the biases in the absolute energy calibration of the LAr calorimeter 
(important only for the {\bf B0} configuration), 

\item the systematic shifts and resolution biases in the reconstructed 
azimuthal angles of particles at the interaction vertices,  

\end{itemize}
to the overall measurement error is significantly smaller that the  
contribution of the luminosity detector effects\footnote{The performance 
precision targets for the host detector can be relaxed by about an order of 
magnitude with respect to those necessary for the precision  measurements of 
the parameters of the electroweak models such as e.g. the mass of the W bosons 
\cite{krasnySMparameters}.}. 

The luminosity detector systematic errors were determined by simulating the 
full data selection and  measurement chain with the biases introduced on 
one-by-one basis. The goal of these simulations was to quantify the impact of  
each of the luminosity detector systematic effects on the final systematic 
uncertainty of the measured luminosity. The results of these simulations can be 
summarised as follows: 
\begin{itemize} 

\item the effect of the luminosity detector misplacement by $0.5$~cm with 
respect to the nominal $z$-collision point of the LHC bunches translates into a 
0.3\% luminosity bias,

\item the effect of decentering of the luminosity detector with respect to the 
beam axis ($x = y = 0$) by $1$~mm translates into a 0.8\% luminosity bias,  

\item the effect of the relative $\phi$-tilt of the luminosity detector planes 
with respect to each other by 1~mrad translates into a 0.1\% luminosity bias,
   
\item the effect of misjudgement of the length of the LHC bunches by 1~cm around 
the central value of 7.5~cm translates into a 0.6\% luminosity bias,  

\item the effect of 0.1\% uncertainty on the value of magnetic field in the 
volume of the host detector tracker and in the fiducial volume of the luminosity 
detector translates into a 0.4\% luminosity bias. 

\end{itemize}
These results show that already for the initial geometrical survey of the 
luminosity detector, before applying the alignment corrections deduced form the 
monitoring data, these contributions are below a 1\% level. Ultimately the 
luminosity detector contribution to the overall measurement uncertainty are 
expected to be driven by the monitoring precision of the length of the LHC 
bunches. The corresponding biases are expected to be smaller than 1\%, provided 
that the length of the LHC bunches is  controlled with a 10\% precision. 

\subsection{Efficiencies} 

The systematic errors of the efficiencies of the electron/positron 
identification and of  the efficiencies of linking of the luminosity detector 
tracks to the host detector tracks  reflect the purity of the monitoring  
sample of small invariant mass electrons and positions pairs originated from 
photons converted in the beam pipe and from the Dalitz decays of neutral pions. 

A special care have to be taken to understand the dependence of the 
electron selection efficiency on the isolation of the electromagnetic cluster. 
For low and medium luminosity runs this can  be done by pre-selecting only those 
of the minimum bias events which are characterised by a low multiplicity of 
particles traversing the fiducial volume of the luminosity detector. For the 
high luminosity runs the efficiency of electron identification  decreases and 
must  be monitored in the instantaneous luminosity dependent way.
  
The impact of the precision of monitoring of $P^{silent}_{LVL2/EF}(t_i)$ on the 
luminosity measurement systematic uncertainty is expected to be negligible in 
the low luminosity periods  in which the $P^{silent}_{LVL2/EF}$ value is 
approaching 1. For medium luminosities this quantity needs to be precisely 
monitored using the dedicated samples of random events. The monitoring precision 
and the corresponding precision of the luminosity measurement depend only 
on the fraction of the total host detector throughput which can be allocated to 
these events.

In general  systematic errors reflecting the achievable precision of monitoring 
of the selection efficiencies can be properly estimated using only real data 
collected at the LHC. They are not expected to contribute significantly to the 
overall measurement uncertainty for the low luminosity periods. To which extent 
this can be achieved for the medium and high luminosity runs remains to be 
demonstrated. 
  
\subsection{Theoretical Cross Section}

The following source of uncertainties on the theoretical calculations of 
$\sigma_{e+e-}$ were taken into account:
\begin{itemize} 

\item the uncertainty of the elastic form factors of the proton,

\item the uncertainty in inelastic form factors of the proton, in the resonance, 
photo-production, transition and in the deep inelastic regions,

\item the uncertainty in the strong and electromagnetic re-scattering cross 
sections,

\item the uncertainty in the higher order electromagnetic radiative corrections.

\end{itemize} 

The first two sources of systematic errors have been analysed and evaluated in 
our previous paper \cite{first}. It was found that for the $e^+e^-$ pairs in the 
fiducial volume of the luminosity detector fulfilling the requirement: 
$\delta\phi_r^{cut} < 0.01$ the present uncertainty on the elastic and inelastic 
proton form factors translates into a 0.3\% precision of the cross section. The 
size of the re-scattering corrections was analysed and found to be smaller than 
0.1\% in the above kinematic region. At present, the largest contribution to the 
total uncertainty comes from the missing calculation of the higher order 
electromagnetic radiative corrections. Their contribution to the cross sections 
can reach 1\%. There is, however, no other than a technical obstacle in 
calculations of these corrections -- if requested they can be made to the 
precision significantly better than the form factors related uncertainties 
\cite{Skrzypek}. 
 
 \section{Conclusions and Outlook}

It has been demonstrated that the proposed method of the luminosity measurement 
has a large potential to provide the highest achievable precision at hadron
colliders. It is based on the electromagnetic collisions of the beam particles 
in the kinematic regime where they can be treated as point-like leptons and, as 
a consequence, their collision cross sections can be calculated with a precision 
approaching the one achieved at the lepton colliders. 

The rate of a fraction of these collisions which can be selected and 
reconstructed using a dedicated luminosity detector  is large enough 
to deliver better that 1\% precision over the data collected over less than one 
month of the data taking with a nominal solenoid current, and a couple of days 
for the ${\bf B0}$ field configuration. The systematic measurement uncertainties 
can be controlled to a better than 1\% precision, by using parasitically, 
samples of the host detector recorded events. The absolute luminosity 
measurement procedures are insensitive to the modeling of the 
collisions mediated by the strong interaction.  

The proposed method can be directly applied by the LHCb experiment which can 
take a full profit from the sizable Lorentz boost of the lepton-pair rest frame 
allowing to replace the electron-positron pairs by the unlike charge muon pairs 
which radiate less and can be identified more easily. Thus the LHCb and the host 
detector distributions could be cross-normalised to a high precision.  

The method presented in this series of papers can be directly used in the 
$pA$ and $AA$ collision modes of the LHC collider. For these collision modes  
the signal-to-background ratio increases, respectively, by $Z^2$ and $Z^4$ due
to the nucleus charge coherence effects. Moreover, a significant increase of the 
signal rate allows to retain its high statistical precision. 

Last but not the least, the scaling of the collider energy dependent rate of the 
luminosity events can be theoretically controlled with a per-mille precision 
allowing to cross normalise the data taken at variable collision energies.  

This series of papers is only a first step towards its ultimate goal: an  
implementation of the proposed method in the LHC environment. The  preliminary 
studies presented in this series of papers, and the use of the base-line 
detector model, are sufficient as a proof of feasibility of the method. Any 
further steps must be preceded by the acceptation of the method by one of the 
LHC collaboration. Should this happen the concrete detector design using the 
host detector preferred technology would be the next step. A conservative 
approach would be to design a detector for the measurement of the absolute 
luminosity in the dedicated low/medium luminosity periods. An ambitious 
programme must have as a target an upgrade the present base-line detector 
concept such that it can be directly used in the high luminosity period of the 
machine operation. This is anything but easy but the gain in the precision of 
all the LHC measurements makes it worth an effort.


\vspace{2cm}   
\begin{thebibliography}{999}
%

\bibitem{Mangano}
M. Mangano, Motivations and precision targets for an
accurate luminosity determination, an opening talk at the CERN  Lumi-days workshop, 
CERN, 13-14 Jnuary 2011. 
%
\bibitem{vdM} 
ATLAS Collab., ATLAS-CONF-2011-116, 19 August 2011. 
%
\bibitem{krasnySMparameters}
M. W.~Krasny, F.~Fayette, W.~Placzek, A.~Siodmok,
Eur. Phys. J. {\bf C51} (2007) 607 (2007) and hep-ph/0702251, \\
 F.~Fayette, M.W.~Krasny, W.~Placzek, A.~Siodmok,
Eur. Phys. J .{\bf C63} (2009) 33 and arXiv:0812.2571 [hep-ph], \\
M. W.~Krasny, F.~Dydak, F.~Fayette, W.~Placzek, A.~Siodmok,
Eur. Phys. J. {\bf C69} (2010) 379 and arXiv:1004.2597 [hep-ex].
%
\bibitem{krasnyHiggs}
M. W. Krasny, 
Acta Phys.Polon. {\bf B42} (2011) 2133 and  arXiv:1108.6163v1 [hep-ph].
%
\bibitem{krasny_jadach}
M. W.~Krasny, S. ~Jadach, W.~Placzek,
Eur. Phys. J. {\bf C44} (2005) 333 and hep-ph/0503215.
%
\bibitem{first} M. W. Krasny, J. Chwastowski and K. S{\l}owikowski, 
Nucl. Instrum. Meth. {\bf A584}~(2008)~42. 
%
\bibitem{second}
M. W.~Krasny, J.~Chwastowski, A. Cyz, and K.~S{\l}owikowski,
Luminosity Measurement Method for the LHC: The Detector Requirements Studies,
June 2010, arXiv:1006.3858 [physics.ins-det].
%
\bibitem{ATLAS} ATLAS Collab., G. Aad et al., 
J. Inst. {\bf 3} (2008) S08003, 
ATLAS Collab.,
CERN-LHCC-2003-022.
%
\bibitem{D0tracker}
D0 Collab., V.M. Abazov et al., Nucl. Instrum. Meth. {\bf A565} (2006) 463.
%
\bibitem{LPAIR} S. P. Baranov, O. Dunger, H. Shooshtari and J.A.M. Vermaseren, 
LPAIR - A Generator for Lepton Pair Production. 
Proceedings of Physics at HERA, vol. 3, (1992) 1478.
%
%
\bibitem{PYTHIA}  T. Sj\"ostrand, P. Ed\'en, C. Friberg, L. L\"onnblad, 
G. Miu, S. Mrenna and E. Norrbin, Computer Phys. Commun. {\bf 135} (2001) 238.
%
\bibitem{Geant} R. Brun et al., Geant 3.21, CERN Program Library Long Writeup W5013,\\
Geant4 Collab., S. Agostinelli et al., Nucl. Instrum. Meth. {\bf A506} (2003) 250,\\
Geant4 Collab., J. Allison et al., IEEE Trans. Nucl. Science {\bf 53} (2006) 270.
%
\bibitem{ATLAS_LUMI} The ATLAS Collab., G. Aad et al., Eur. Phys. J. C (2011) 
71: 1630.
%
\bibitem{Myers} S. Myers, private communication
%
\bibitem{RomanPot} Proceedings of the Workshop on Fast Timing Detectors:
Electronics, Medical and Particle Physics Applications, November 29 -- 
December 1, 2010, eds. J. Chwastowski, P. Le Du, C. Royon, Acta. Phys. 
Polonica B Proceedings Supplement, vol. 4, no. 1, 2011.
\bibitem{micromegas} Y. Giomataris, Ph. Rebourgeard, J-P. Robert and G. Charpak,
Nucl. Instr. Meth. {\bf A376} (1996) 29.
\bibitem{Skrzypek} M. Skrzypek and S. Jadach, private communication.
%
\bibitem{Charpak} 
Y.Giomataris, G. Charpak., 
Nucl. Instr. Meth.   {\bf A310} (1991) 589.
\end{thebibliography}
\end{document}